\documentclass[11pt,a4paper]{article}
\pdfoutput=1

\usepackage{jcappub}
\usepackage[margin=0.95in]{geometry}

\usepackage[utf8]{inputenc}
\usepackage{graphicx}
\usepackage{verbatim}
\usepackage{epsfig}
\usepackage{subfigure}
\usepackage{color}
\usepackage{multirow}
\usepackage{comment}
\usepackage{slashed}
\usepackage{amsmath}
\usepackage{ulem}
\usepackage{framed}

\usepackage{amssymb,amsmath,amsfonts,mathrsfs}
\usepackage[dvipsnames]{xcolor}
\usepackage{graphicx}
\usepackage{longtable}
\usepackage{verbatim}
\usepackage{color}
\usepackage[mathscr]{euscript}
\usepackage{framed}
\usepackage{mdframed}  

\usepackage{cancel}
\usepackage{color}
\usepackage{hyperref}
\hypersetup{colorlinks, citecolor=bluscuro, linkcolor=black, urlcolor=bluscuro}
\definecolor{rossos}{cmyk}{0,1,1,0.55}
\definecolor{bluscuro}{rgb}{0.15, 0.2, .85}
\definecolor{bluchiaro}{cmyk}{1,.3,0.,0.1}

\numberwithin{equation}{section}
\renewcommand\theequation{\arabic{section}.\arabic{equation}}

\usepackage{tikz}

\usepackage{tcolorbox}
\newcommand{\nn}{\nonumber}

\newcommand{\lp }[0]{\left (}
\newcommand{\rp }[0]{\right )}
\newcommand{\llp }[0]{\left [}
\newcommand{\rrp }[0]{\right ]}

\newcommand{\med}[1]{\langle #1\rangle}

\newcommand{\HH}{{\cal H}}

\newcommand{\vk}{\vec{k}}
\newcommand{\vx}{\vec{x}}
\newcommand{\vy}{\vec{y}}

\def\PBH{\text{\tiny PBH}}
\def\pk{\text{\tiny pk}}
\def\CMC{\text{\tiny cmc}}
\def\H{\text{\tiny H}}

\newcommand{\be}{\begin{equation}\begin{aligned}}
\newcommand{\ee}{\end{aligned}\end{equation}}

\newcommand{\bbe}{\begin{align}}
\newcommand{\eee}{\end{align}}

\newcommand{\bea}{\begin{eqnarray}}
\newcommand{\eea}{\end{eqnarray}}

\def\beq{\begin{equation}}
\def\eeq{\end{equation}}

\def\d{{\rm d}}

\def\vk{{\vec{k}}}

\def\beqa{\begin{eqnarray}}

	\def\eeqa{\end{eqnarray}}
	
\def\lsim{\mathrel{\rlap{\lower4pt\hbox{\hskip0.5pt$\sim$}}
		\raise1pt\hbox{$<$}}}         %less than or approx. symbol
\def\gsim{\mathrel{\rlap{\lower4pt\hbox{\hskip0.5pt$\sim$}}
		\raise1pt\hbox{$>$}}}         %greater than or approx. symbol

\def\d{{\rm d}}

\def\pk{{\text{\tiny pk}}}

\def\d{{\rm d}}

\def\PBH{\text{\tiny PBH}}

\def\eeqa{\end{eqnarray}}

\def\bq{\begin{quote}}
\def\eq{\end{quote}}

 at 10truept
\newcommand{\arXiv}[2]{\href{http://arxiv.org/pdf/#1}{{\tt [#2/#1]}}}
\newcommand{\arXivold}[1]{\href{http://arxiv.org/pdf/#1}{{\tt [#1]}}}

\def\eeqa{\end{eqnarray}}
\def\lsim{\mathrel{\rlap{\lower4pt\hbox{\hskip0.5pt$\sim$}}
    \raise1pt\hbox{$<$}}}         %less than or approx. symbol
\def\gsim{\mathrel{\rlap{\lower4pt\hbox{\hskip0.5pt$\sim$}}
    \raise1pt\hbox{$>$}}}         %greater than or approx. symbol

\title{The Initial Spin Probability Distribution of Primordial Black Holes}
\author[a]{V. De Luca,}
\author[b]{V. Desjacques,}
\author[a]{G. Franciolini,}
\author[a]{A. Malhotra,}
\author[a,c]{\textcolor{White}{---------} A. Riotto}

\affiliation[a]{
	Department of Theoretical Physics and Center for Astroparticle Physics (CAP) \\
			24 quai E. Ansermet, CH-1211 Geneva 4, Switzerland}
\affiliation[b]{Physics department and Asher Space Science Institute, Technion, Haifa 3200003, Israel}
\affiliation[c]{CERN, Theoretical Physics Department, Geneva, Switzerland}

\abstract{We study the spin of primordial black holes produced by the collapse of large inhomogeneities in the early universe. Since such primordial black holes originate from peaks, that is, from   maxima of the local overdensity, we resort to peak theory to obtain the probability  distribution of the spin at formation.  We show that the spin is a first-order effect in perturbation theory: it results from the action of first-order tidal gravitational fields  generating  first-order torques upon horizon-crossing,  and from the asphericity of the  collapsing object. Assuming an ellipsoidal shape, the typical value of the dimensionless parameter  $a_{\rm s}=S/G_N M^2$, where $S$ is the spin and $M$ is the mass of the primordial black hole, is about $\sigma_\delta\sqrt{1-\gamma^2}/2\pi$. Here, $\sigma^2_\delta$ is the variance of the overdensity at horizon crossing
and the parameter $\gamma$ is  a measure of the width of the power spectrum giving rise to primordial black holes. One has $\gamma=1$ for monochromatic spectra. For these narrow spectra, the suppression arises because the velocity shear, which is strongly correlated with the inertia tensor, tends to align with the principal axis frame of the collapsing object. Typical values of $a_{\rm s}$ are at the percent level. }
\emailAdd{valerio.deluca@unige.ch}
\emailAdd{dvince@physics.technion.ac.il}
\emailAdd{gabriele.franciolini@unige.ch}
\emailAdd{ameek.malhotra@etu.unige.ch}
\emailAdd{antonio.riotto@unige.ch}
\begin{document}

\maketitle
\flushbottom

\section{Introduction}
The physics of black holes  has attracted a lot of attention since the recent discovery of the gravitational waves generated by the merger of rather massive black holes \cite{ligo}.  This interest is also motivated by the fact that PBHs might compose a fraction of the dark matter of the universe \cite{bird} (for a review, see Ref. \cite{sasaki}). 

One particularly interesting property of the black holes is their spin.  Larger integrated fluxes of gravitational waves are obtained for larger  spins as they lead to   smaller ISCO separations in binaries and to longer inspiral phases. The  observations of gravitational waves sourced emitted during the merging of two massive black holes may be exploited to  measure the spin of the final state, as well as  the orbital projection of the effective spin of the black hole binary
\be
\chi_{\rm eff}=\frac{M_1\vec{S}_1+M_2\vec{S}_2}{M_1+M_2}\cdot \hat{L}.
\ee
Here, $\vec{S}_1$ and $\vec{S}_2$ are the spins of the initial merging black holes in the binary and $\vec{L}$ their initial angular momentum.
Currently, the handful number of gravitational wave measurements indicates that $\chi_{\rm eff}$  is compatible with zero \cite{ligospin}.

There are currently two  major astrophysical models of black hole spin distributions. One possibility is that the final spin of the binary system and the angular momentum are aligned, as expected for black hole binaries formed in  a shared  envelope evolution within  galactic fields \cite{bl}. Alternatively, the initial spins have an  isotropic distribution as expected in binaries originated  in globular or stellar clusters in the proximity of  active galactic nuclei \cite{ac}. 
There can be another mechanism which  produces small effective spins.  This is likely if the  black holes are of primordial origin\footnote{Massive black holes of stellar origin may also be  born with small natal spins \cite{Belczynski:2017gds}.}, hence the name Primordial Black Holes (PBHs),  and they are produced by the collapse of sizeable
inhomogeneities in the early universe. This case is particularly interesting because numerical relativity  simulations provide evidence that  two black holes with initially small spin
merge into a bigger black hole of mass $M_f$ possessing a  final spin $S_f$ of the order of \cite{ba}
\be
a_f=\frac{S_f}{G_NM_f^2}\simeq 0.69-0.56\left(\frac{M_1-M_2}{M_1+M_2}\right)^2,
\ee
where $G_N$ is Newton's constant.
This result is robust to the details of the evolution, and to the choice of the initial binary orbital angular momentum in the case of quasi-circular orbits. This prediction for $a_f$ is compatible with current measurements. 

PBHs can be generated in the early Universe through the   collapse of large enough  density perturbations. In particular, inflation can lead to PBH production when the comoving curvature perturbation ${\cal R}$  is enhanced on small scales (much smaller than the scales accessible to the CMB). 
The process of   reheating  after inflation transfers these sizeable  inhomogeneities to  radiation, and  PBHs form when these extreme perturbations re-enter  the horizon \cite{sasaki}. 

While the PBH spins are isotropically distributed at the PBH formation epoch, the probability distribution of the spin amplitude at PBH formation time is not known. This will be the focus  of the present  work.
We shall study the PBH spin probability distribution at  formation time. Such a discussion is, to the best of our knowledge, absent in the literature. To proceed, we will use standard results in peak theory  \cite{Bardeen:1985tr} which is the appropriate framework since PBHs originate from peaks or, more precisely, from local maxima of the radiation density distribution. 

We will show that the zeroth-order anisotropy of the collapsing object giving rise to a PBH is coupled to the first-order tidal gravitational field and generate a first-order torque. This happens if one takes into account the fact that the density profile around peaks generally deviates from spherical symmetry. This result bears a strong similarity with the well-known ``tidal-torque theory'' in large-scale structure \cite{doroshkevich,white,hoffman,barnesefstathiou,torques,SME}. For instance, as demonstrated in Ref. \cite{white}, the classical result of Peebles \cite{peebles} that the spin of halos is generated at second-order is a consequence of choosing a Lagrangian sphere to describe the collapsing object. In our case, the shape of the Lagrangian region which collapses to form a PBH is imprinted when the characteristic scale of the perturbation is superhorizon.  

Our findings indicate  that the spin does not vanish at first-order in perturbation theory because, in the homogeneous ellipsoid approximation adopted throughout this work (the gravitational collapse of such ellipsoidal perturbations has been investigated extensively in the large scale structure literature \cite{lyndenbell,linmestelshu,fujimoto,zeldovich,icke,whitesilk,barrowsilk,hb96,bondmyers,SMT,VD}), the lengths of the principal axes of the inertia tensor are different. However, another necessary condition to generate a non-zero spin is the non-vanishing of the off-diagonal components of the velocity shear. In other words, the ellipsoidal perturbation tracing the peak from which the PBH originates must have its inertia tensor misaligned with that of the velocity shear. This does not happen when the perturbations are on superhorizon scales. Notwithstanding, 
a torque is generated once the perturbation re-enters the horizon until it decouples from the background, i.e. until turnaround. This generates a small, albeit non-zero spin. 
After turnaround, we speculate that the angular momentum remains constant, as seen for the collapse of non-relativistic dust \cite{peebles,white,barnesefstathiou}. However, this issue should be explored further with numerical simulations.

We will show that the typical value of the  dimensionless Kerr spin parameter $a_{\rm s}=S/G_N M^2$ (where $M$ is the mass of the PBH) is
\be
a_{\rm s}=\frac{1}{2\pi} \sigma_\delta\sqrt{1-\gamma^2}\sim 10^{-2}\sqrt{1-\gamma^2},  
\ee 
where $\sigma^2_\delta$ is the variance of the overdensity at horizon crossing
and $\gamma$ measures how narrow is the power spectrum giving rise to the PBHs. Very narrow power spectra have $\gamma\simeq 1$ (a monochromatic power spectrum, or Dirac delta, has exactly $\gamma=1$). Hence, the smaller the PBH spin,  the narrower the spectrum.
As we shall see, the suppression factor due to the dependence on $\gamma$ arises because, as $\gamma$ approaches unity, the velocity shear tends to be more strongly aligned with the inertia tensor.

The  conditional spin probability $a_{\rm s} P(a_{\rm s}|\nu)$ (where $\nu=\delta/\sigma_\delta$ parametrises the height of the peak) exhibits some interesting features: for higher peaks, the PBH spin shifts towards smaller values and, once the height of the  peak is held fixed, slightly smaller values of the spins are obtained the steeper the power spectrum is.

Our paper is organised as follows. In Section 2, we start from peak theory to characterise the spin at first- and second-order in perturbation theory. In Section 3, we
study the spin on superhorizon scales, while Section 4 is dedicated to the subhorizon treatment. Section 5 describes the  PBH dimensionless Kerr parameter at formation time. Section 6
contains technical details of the computation of the spin probability distribution. Section 7 illustrates how the  PBH spin is correlated with the shape of the power spectrum. Section 8 discusses the impact of the spin onto the PBH abundance. Finally, Section 9 summarizes our conclusions.

\section{The spin of PBHs as local density maxima}
\label{statistics}
The definition of spin or angular momentum in general relativity is a delicate issue, see for instance  Ref. \cite{ang} for a detailed discussion and more recently Ref. \cite{nor}.  
The angular momentum represents a conserved quantity originated from rotational  invariance. Naively, one might define it through the (3+1)-formalism as the conserved quantity associated with rotations at spatial infinity  
and to  asymptotically flat spacetime observers. This would be the generalisation of  the ADM momentum obtained by replacing the translation Killing vectors  at spatial infinity by the corresponding  rotational Killing vectors.  
The problem with this definition  is that the resulting quantity does not transform as a vector under the change of the coordinates preserving the asymptotic properties at infinity. The issue arises from the existence of the so-called supertranslations at infinity \cite{york}.
However, if the spacetime possesses  some symmetries, global quantities may be defined which are   coordinate-independent through the technique introduced by Komar \cite{komar} which amounts to   taking flux integrals of the derivative of the Killing vector associated with the symmetry over closed two-surfaces surrounding the matter sources. These  quantities are conserved in the sense that they do not depend on the choice of the surface provided that all the matter is included in it. One can therefore define the amplitude of the  Komar angular momentum  on a given time-slicing $\Sigma$  through the Komar integral \cite{ang}
\be
S(\Sigma)=\frac{1}{16\pi G_N}\int_{\partial\Sigma}{\rm d}S_{\mu\nu}D^\mu\phi^\nu,
\ee
where $\phi^\nu$ is the  rotational Killing vector of the asymptotic flat metric. Using Gauss' law and Einstein equations, one can rewrite this expression as
\be
S(\Sigma)=\int_{\Sigma}{\rm d}S_{\mu}J^\mu(\phi),
\ee
where $J^\mu(\phi)=T^\mu_{\,\,\nu}\phi^\nu-T\phi^\mu/2$ is expressed in terms of the  energy-momentum tensor. For a relativistic perfect fluid one gets
\be
 S(\Sigma)=\int_{\Sigma}\d V\, T^0_{\,\,\mu}\phi^\mu= \frac{4}{3}\int_\Sigma\d V \rho\, \vec{v}\cdot \vec{\phi},
\ee
where  $\vec v$ is the velocity field and   $\rho$ is the density field.
In the following we will be concerned with the spin of perturbations collapsing around a local maxima of the overdensity. Therefore, expanding around the peak  located at $\vec{x}_\pk$, one obtains
\be
\label{jj}
S_i = \frac{4}{3} a^4(\eta) \epsilon_{ijk}\int \d^3x\,\sqrt{\gamma}\, \rho(\vec{x},\eta)(x- x_\pk)^j (v- v_\pk)^k,
\ee
where $\vec x$ is a comoving coordinate and we have gone to conformal time. The presence of the velocity, which is a first-order quantity in cosmological perturbation theory, makes the spin  at least a first-order quantity. Of course, by expanding the energy density $\rho(\vec x,\eta)$ in perturbation theory one would obtain higher-orders contribution to the spin.
However, in the rest of the paper we will be concerned with the spin of PBHs only at first-order.

\subsection{First-order description of the PBH spin}
The 3-dimensional volume $V_{\rm e}$ over which we perform the integral Eq. (\ref{jj}) could be significantly aspherical. To proceed further, we associate the PBHs to high peaks of the overdensity and use the triaxial ellipsoid approximation to prescribe the volume $V_{\rm e}$.
We then use the standard results of peak theory and study the statistics of local maxima \cite{Bardeen:1985tr}.
Expanding the overdensity around the peak up to second-order (in the derivatives), we obtain\footnote{We choose to work with the density contrast instead of the curvature perturbation to avoid un-physical IR effects arising from super-horizon modes, see for example Ref.~\cite{sasaki1}.}
\be
\delta(\vx) = \frac{\delta \rho(\vx)}{\bar\rho} \simeq \delta_\pk + \frac{1}{2}\zeta_{ij}(x- x_\pk)^i (x- x_\pk)^j > f \delta_\pk, \qquad  \zeta_{ij} = \left.\frac{\partial^2 \delta}{ \partial x^i \partial x^j}\right|_\pk,
\ee
where the first gradient term is zero as we are dealing with maxima of the overdensity. The parameter $f$ is set by  the criterion to form a PBH, that is that the  matter nearby the peak  be above a certain overdensity threshold.
Performing a rotation of the coordinate axes to be aligned with the principal axes of length $\lambda_i$ of the constant-overdensity ellipsoids gives
\be
\label{alambda}
\delta \simeq \delta_\pk -\frac{1}{2}\sigma_{\zeta} \sum_{1}^{3}\lambda_i (x^i- x_\pk^i)^2 .
\ee
Here, $\sigma_{\zeta}$ is the characteristic rms variance of the components of $\zeta_{ij}$. We purposely avoid the spectral moment notation of Ref. \cite{Bardeen:1985tr} to remain as general as possible. As we will see later, the velocity shear and the density perturbations which contribute to the linear tidal-torque effect   can expressed through the familiar spectral moments of the density power spectrum.

Eq. (\ref{alambda}) can be rearranged as
\be
\frac{2(\delta - \delta_\pk)}{\sigma_{\zeta}} = - \sum_{1}^{3}\lambda_i (x^i- x_\pk^i)^2.
\ee
Focusing on the approximately ellipsoidal surface $\delta = f \delta_\pk$ and defining the height of the peak $\nu$ in units of the rms overdensity $\sigma_\delta$ as $\nu = \delta_\pk/\sigma_\delta$, leads to 
\be
2\frac{\sigma_\delta}{\sigma_{\zeta}}(1-f)\nu =  \sum_{1}^{3}\lambda_i (x^i- x_\pk^i)^2,
\ee
whose solutions define the boundary of the integration volume in the spin integral. They  are the principal semi-axes of such an ellipsoid and are given by\footnote{Notice that  $a^2_i$, being ratios of first-order quantities,  are  not  small quantities and this  is the ultimate reason why
tidal-torque theory, when applied to dark matter  halos, predicts a much larger spin
parameter although one starts from a density fluctuation at recombination
which is of order $(10^{-4} \div 10^{-3})$.}
\be
\label{semiaxes}
a_i^{2} = 2\frac{\sigma_\delta}{\sigma_{\zeta}}\frac{(1-f)}{\lambda_i}\nu.
\ee
Notice that  in the limit of large $\nu$ one obtains the following useful expressions \cite{Bardeen:1985tr}
\be
\lambda_i=\frac{\gamma\nu}{3}\left(1+\epsilon_i\right),\,\,\,\epsilon_i={\cal O}\left(\frac{1}{\gamma\nu}\right)\,\,\,\,{\rm and}\,\,\,\,a_i^{2}\sim 6\frac{\sigma_\delta^2}{\sigma_{\times}^2},
\ee
where
\be
\gamma= \frac{\sigma_{\times}^2}{\sigma_\delta\sigma_{\zeta}}
\ee
and $\sigma_{\times}^2$ is the characteristic cross-correlation between $\delta$ and $\zeta_{ij}$.
Eq. (\ref{semiaxes}) shows that the principal semi-axes behaves like $a_i^2  \sim\sigma_\delta\nu/\sigma_\zeta|\zeta_{ij}|\sim R_*^2$, where $R_*$ is a characteristic scale to be defined below. Furthermore, the difference $|\lambda_i-\lambda_j|$ is of order unity because the ``ellipticity'' $\epsilon_i$ scales like $1/(\gamma\nu)$.

Next, we first   expand the velocity field around the peak 
\be
(v- v_\pk)^k = v_l^k (x- x_\pk)^l, \qquad v_l^k = \frac{\partial v^k}{\partial x^l}\bigg|_\pk,
\ee
to get
\begin{eqnarray}
\label{rc}
S_i &=&  \frac{4}{3}a^4(\eta) \epsilon_{ijk} \overline{\rho}_{\rm rad}(\eta) \int_{V_{\rm e}} \d^3x (x- x_\pk)^j(v- v_\pk)^k\nonumber\\
&=&\frac{4}{3}a^4(\eta) \epsilon_{ijk} \overline{\rho}_{\rm rad}(\eta) v_l^k\int_{V_{\rm e}} \d^3x (x- x_\pk)^j(x- x_\pk)^l.
\end{eqnarray}
In this expression we have expanded at zeroth-order the density field $\rho(\vec x, t)$, and  defined the average density of the universe at a given time $\eta$, $\overline{\rho}_{\rm rad}(\eta)$, which we suppose to be dominated by radiation. 

The volume $V_{\rm e}$ over which the integral is performed generally is a function of time. Internal and external tidal forces will gradually deform it until the whole perturbation collapses to a black hole. However, since the characteristic (comoving) size of the perturbation is $k_\H^{-1}\sim M_{\PBH}^{1/2}$, the deformation becomes significant only once the perturbation has re-entered the horizon and decoupled from the background, i.e. after turnaround. Before turnaround, we can approximate the perturbation as a rigid body of volume $V_{\rm e}$ whose boundary is the isodensity surface $\delta=f\delta_\pk$.
The value of $f<1$ is chosen such that $f\delta_\pk$ equates the critical threshold $\delta^{\rm c}$ for PBH collapse. This critical density  depends on the choice of gauge (see below).

The spin can be found by performing the integration over the ellipsoidal volume $V_{\rm e}$ appearing in its definition as
\be\label{spin-1-ord}
S_i = \frac{4}{3}a^4(\eta) \epsilon_{ijk} \overline{\rho}_{\rm rad}(\eta) v_l^k\int_{V_{\rm e}} \d^3x \  x^j x^l =
 \frac{4}{3}a^4(\eta) \epsilon_{ijk}  \overline{\rho}_{\rm rad}(\eta)g_v(\eta)\widetilde{v}_{kl}\int_{V_{\rm e}} \d^3x \  x^j x^l,  
\ee
where,  without loss of generality,  we have chosen $\vec{x}_\pk = 0$  and we  have factorised out the time-dependence of $v_l^k$ as 
\be
\label{eq:deftildev}
v_l^k(\eta)=g_v(\eta)\widetilde{v}_{l}^k.
\ee
Eq. (\ref{eq:deftildev}) defines the normalised velocity shear $\widetilde v_{kl}$ (where the scale factor can be normalised to unity today). 
We perform  the integration by doing a change of coordinates to ellipsoidal coordinates of the form
\be\label{changecord}
\int_{V_{\rm e}}   \d^3x =a_1 a_2 a_3  \int _0 ^1 r^2 \d r \int_0 ^{2 \pi}\d \phi \int_0 ^{ \pi}   \d \theta\,\sin\theta  
\ee
and identifying 
\be
x_1 = a_1 r \cos \phi \sin \theta, \quad x_2 =a_2  r \sin \phi \sin \theta, \quad x_3=a_3 r \cos \theta.
\ee
Thus we get
\be
\int_{V_{\rm e}} \d^3x\,  x^j x^l  = 
\frac{4 \pi}{15} a_1 a_2 a_3  \begin{bmatrix}
	a_1^2&0 & 0
	\\
	0 &a_2^2 &0 
	\\
	0&0 &a_3^2
\end{bmatrix}_{jl}.
\ee
Finally, Eq.~\eqref{spin-1-ord} can be written as \cite{torques}
\begin{tcolorbox}[colframe=white,arc=0pt]
\vspace{-.15cm}
\be
\label{f}
\vec S^{(1)} = \frac{16\pi}{45} a^4(\eta) \overline{\rho}_{\rm rad}(\eta)g_v(\eta)  a_1 a_2 a_3 ([a_2^2-a_3^2]\widetilde{v}_{23}, [a_3^2-a_1^2]\widetilde{v}_{13},[a_1^2-a_2^2]\widetilde{v}_{12}).
\ee
\end{tcolorbox}
\noindent
This shows that the spin does not vanish at first-order only if {\it i}) the lengths of the semi-axis are different and {\it ii}) the off-diagonal components of the velocity shear are non-zero, and misaligned with the   inertia tensor.
For later use, using  Eq. \eqref{semiaxes} we can recast the spin in the form
\begin{eqnarray}
\label{lll}
\vec S^{(1)}&=&\frac{4}{3} \frac{16 \sqrt{2}\pi}{15}  a^4(\eta) \overline{\rho}_{\rm rad}(\eta)g_v(\eta) \lp \frac{\sigma_\delta}{\sigma_{\zeta}} \rp ^{5/2} \frac{(1-f)^{5/2} \nu^{5/2}}{\sqrt{\lambda_1 \lambda_2 \lambda_3}} 
\lp -\alpha_1 \widetilde{v}_{23}, \alpha_2 \widetilde{v}_{13},
-\alpha_3 \widetilde{v}_{12} \rp 
 \nonumber \\
&=&  \bigg[ \frac{4}{3}a^4(\eta) \overline{\rho}_{\rm rad}(\eta)g_v(\eta)(1-f)^{5/2} R_*^5 \bigg] \frac{16 \sqrt{2}\pi}{135\sqrt{3}}\left(\frac{\nu}{\gamma}\right)^{\frac{5}{2}}\frac{1}{\sqrt{\Lambda}}\lp -\alpha_1 \widetilde{v}_{23}, \alpha_2 \widetilde{v}_{13},
-\alpha_3 \widetilde{v}_{12} \rp,\nonumber\\
&&
\end{eqnarray}
where
\be
\alpha_1 = \frac{1}{\lambda_3}-\frac{1}{\lambda_2}, \ \ \alpha_2 = \frac{1}{\lambda_3}-\frac{1}{\lambda_1}, \ \ \alpha_3 = \frac{1}{\lambda_2}-\frac{1}{\lambda_1},
\ee
are defined such that $\alpha_i \geq 0$, with $\alpha_2 \geq \alpha_1, \alpha_3$ due to the labelling choice $\lambda_1> \lambda_2> \lambda_3$.
Here, 
\be
\Lambda = \lambda_1 \lambda_2 \lambda_3, \,\,\,\, R_*=\sqrt{3}\frac{\sigma_{\times}}{\sigma_{\zeta}}.
\ee
The term inside the square brackets in the last line of Eq. (\ref{lll}) identifies the time-dependent part of the spin, that we will label as the reference spin
\be
S_{\rm ref}(\eta) = \frac{4}{3} a^4(\eta) \overline{\rho}_{\rm rad}(\eta)g_v(\eta)R_*^5 (1-f)^{5/2}. 
\ee
It is the same for all peaks, while the information on the shape and the height of the peak is carried by the remaining term. The magnitude of the dimensionless spin 
can therefore be written as 
\be
\label{S}
 S^{(1)}(\eta) = S_{\rm ref}(\eta) s^{(1)}_{\rm e}.
\ee
While $S_{\rm ref}(\eta)$  is common to all   peaks,  $s^{(1)}_{\rm e}$ depends on the shape and the height of the peaks.
This parametrisation isolates the time-dependence in such a way that its time derivative may be thought of as the torque acting on matter in the neighbourhood of local density maxima.

As we can see, since $\alpha_i\sim |\lambda_j-\lambda_k|/\lambda_j\lambda_k \sim 1/(\gamma\nu)^2$, $s^{(1)}_{\rm e}$ genuinely is a quantity of order ${\cal O}(1)$ (up to multiplicative factors of $\gamma$ and $\nu$). The only first order quantity appearing in this calculation thus is the velocity shear $v_l^k$. Therefore, the spin $S^{(1)}(\eta)$ is truly first-order.

\subsection{Second-order description of the PBH spin}
For the sake of completeness, we present here   the second-order expansion obtained taking one more term when perturbing the energy density
\be
\rho (\vec x,\eta ) = \overline{\rho}_{\rm rad} (\eta) + \delta \rho =\overline{\rho}_{\rm rad} (\eta) \left( 1+ \delta \right )=
\overline{\rho}_{\rm rad} (\eta) \llp  1+ \delta_\pk + \frac{1}{2}\zeta_{ij} (x-x_\pk)^i  (x-x_\pk)^j \rrp 
\\
=\overline{\rho}_{\rm rad}  (\eta) \lp  1+ \delta_\pk \rp +\overline{\rho}_{\rm rad} (\eta) \llp \frac{1}{2}\zeta_{ij} (x-x_\pk)^i  (x-x_\pk)^j \rrp.
\ee
This corresponds to consider the second-order effects in the spin
\be
S_i = \frac{4}{3} a(\eta)^4 \epsilon_{ijk}\overline{\rho}_{\rm rad}(\eta) v_l^k \llp \lp 1+ \delta_\pk \rp\int_{V_{\rm e}}  \d^3x\,  x^j x^l  + 
\frac{1}{2} \zeta_{mn} \int_{V_{\rm e}}\d^3x \,  x^j x^l x^m x^n \ \rrp,
\ee
where $V_{\rm e}$  is the ellipsoidal volume.
We perform  the integrations by doing a change of coordinates to ellipsoidal coordinates as in Eq.~\eqref{changecord}
and, defining $I^{jlmn} =\int_{V_{\rm e}}\d^3x\, x^j x^l x^m x^n$, one finds
\be
I^{1111}&=\frac{4 \pi}{35}a_1^5 a_2 a_3, \quad I^{2222}=\frac{4 \pi}{35}a_1 a_2^5 a_3,
\quad I^{3333}=\frac{4 \pi}{35}a_1 a_2 a_3^5,
\\
 I^{2211}&=I^{1122}=I^{2121}=I^{2112}=I^{1221}=I^{1212}=\frac{4 \pi}{105}a_1^3 a_2^3 a_3,
\\
 I^{3311}&=I^{1133}=I^{3131}=I^{3113}=I^{1331}=I^{1313}=\frac{4 \pi}{105}a_1^3 a_2 a_3^3,
\\
I^{3322}&=I^{2233}=I^{2332}=I^{2323}=I^{3232}=I^{3223}=\frac{4 \pi}{105}a_1 a_2^3 a_3^3,
\ee
where all the other components are zero. The spin up to second-order is therefore given by 
\be
\vec S =\vec S^{(1)}(1+\delta_\pk)+\vec S^{(2)},
\ee
where the components of $\vec S^{(2)}$ are
	\be
        \label{eq:spinsecondorder}
	S^{(2)} _1& =  \frac{16\pi}{90} a^4(\eta)\overline{\rho}_{\rm rad}(\eta)  g_v(\eta)  a_1 a_2 a_3 \cdot \frac{1}{7}
	\llp 
	3 a_2^4 \widetilde v_{23} \zeta_{22} - 3 a_3^4 \widetilde v_{23} \zeta_{33}
	+
	a_1^2  a_2^2 (\widetilde v_{23} \zeta_{11} + 2 \widetilde v_{13} \zeta_{12}) 
	\right .
	\\
	& \left .  
	- 	a_1^2  a_3^2 ( \widetilde v_{23} \zeta_{11} + 2 \widetilde v_{12} \zeta_{13})
	+ a_2^2 a_3^2 (-\widetilde v_{23} \zeta_{22} - 2 \widetilde v_{22} \zeta_{23} + 2 \widetilde v_{33} \zeta_{23} + 
	\widetilde v_{23} \zeta_{33})    \rrp ,
	\\
	S^{(2)} _2& =  \frac{16\pi}{90} a^4(\eta)\overline{\rho}_{\rm rad}(\eta)  g_v(\eta)  a_1 a_2 a_3 \cdot \frac{1}{7}
	\llp  -3 a_1^4 \widetilde v_{13} \zeta_{11} 
	+ 
	3 a_3^4 \widetilde v_{13} \zeta_{33}
	-a_1^2 a_2^2 (2 \widetilde v_{23} \zeta_{12} + \widetilde v_{13} \zeta_{22})
	\right.    \\
	&  \left .  
		+ 
	a_2^2 a_3^2 (\widetilde v_{13} \zeta_{22} + 2 \widetilde v_{12} \zeta_{23})  
	 + 
	a_1^2 a_3^2 (\widetilde v_{13} \zeta_{11} + 2 \widetilde v_{11} \zeta_{13} - 2 \widetilde v_{33} \zeta_{13} - 
	\widetilde v_{13} \zeta_{33})
	\rrp ,
	\\
	S^{(2)} _3 & =  \frac{16\pi}{90} a^4(\eta)\overline{\rho}_{\rm rad}(\eta)  g_v(\eta)  a_1 a_2 a_3 \cdot \frac{1}{7}
	\llp  3 a_1^4 \widetilde v_{12} \zeta_{11} 
	- 
	3 a_2^4 \widetilde v_{12} \zeta_{22}+ 
	a_1^2 a_2^2 (-\widetilde v_{12} \zeta_{11} - 2 \widetilde v_{11} \zeta_{12} \right . \\
        & \left . + 2\widetilde  v_{22} \zeta_{12} + \widetilde  v_{12} \zeta_{22}) 
        + a_1^2 a_3^2 (2 \widetilde v_{23} \zeta_{13} + \widetilde v_{12} \zeta_{33})
	- a_2^2 a_3^2 (2 \widetilde v_{13} \zeta_{23} +\widetilde  v_{12} \zeta_{33})
	\rrp .
	\ee
We will now discuss the two relevant regimes, super- and sub-horizon.

\section{The PBH spin before horizon crossing}
\label{sec:beforecrossing}
In the early radiation-dominated Universe, PBHs are generated when  highly overdense regions gravitationally collapse directly into a black hole. Before collapse, the comoving sizes of such  regions are   larger than the horizon length and the  separate universe approach can be applied \cite{harada}.
We can  therefore expand at  leading order in spatial gradients of the various observables, e.g. the overdensity.  At this stage, one should  fix  the slicing and the threading of the spacetime manifold.  The so-called constant mean curvature slicing (CMC, sometimes dubbed the uniform Hubble slicing) seems appropriate as it has been adopted to perform numerical relativity simulations to describe  the formation of PBHs and to calculate the  threshold for PBH formation \cite{shibata}. In this slicing the equations for the lapse function and the gravitational potentials are similar to those one encounters
in the maximal slicing gauge (in which the total curvature vanishes) in the asymptotically flat spacetime. 
 We will provide further considerations about this gauge in section 4.

In the CMC slicing, the overdensity
turns out to be \cite{harada}
 \be
 \delta(\vec x,\eta)=-\frac{4}{3}e^{-5\mathcal{R}(\vec x)/2}\frac{\nabla^2e^{\mathcal{R}(\vec x)/2}}{{\cal H}^2},
 \ee
where $\mathcal{R}(\vec x)$ is the comoving curvature perturbation. This expression is fully non-perturbative, being the only expansion in gradients. Notice that the coefficient is 3/2 times larger than the one  in the comoving slicing (CG), as it is the corresponding threshold for PBH formation, i.e. $\delta^{\rm c}_{\CMC} =3 \delta^{\rm c}_{\text{\tiny CG}} /2$. We are going to consider, as a reference value,  $\delta^{\rm c}_{\text{\tiny CG}} =0.45$,  and therefore $\delta^{\rm c}_{\CMC} =0.675$ \cite{harada}, neglecting the shape dependence of the threshold \cite{haradath,mg,musco}.

As the universe expands,  the overdensity  grows.  Regions where it  becomes of order unity eventually stops expanding and collapse. This happens when the comoving scale of such a region becomes of the order of the  horizon scale. Even though the gradient expansion approximation  breaks down, it has been used to obtain  an acceptable criterion for the PBH  formation, as confirmed by  nonlinear numerical studies \cite{sasaki}, and we will follow the same strategy here. The spin grows until the system decouples from the background expansion and  torques are reduced.

In the threading where the pertubed ${}_{0i}$ component of the metric vanishes, the velocity in the long wavelength regime is given by \cite{harada}
\be
v^i(\vec x,\eta)=\frac{1}{12 {\cal H}}\partial^i\delta(\vec x,\eta ).
\ee
The velocity therefore scales like $a^3$ on superhorizon scales.

Since the velocity is proportional to the gradient of the overdensity, from Eq. (\ref{alambda}) one immediately concludes that the off-diagonal entries of the matrix $\widetilde{v}_{ij}$ are vanishing. Therefore, before horizon-crossing the first-order PBH spin $\vec{S}^{(1)}$ is zero. The physical reason is that, in the  coordinate frame aligned with
the principal axis frame of the inertia tensor, the velocity shear $v_l^k$ is aligned with the inertia tensor and no spin can be generated independently of the deviation from sphericity of the collapsing region and at any order in perturbation theory in the comoving curvature perturbation $\mathcal{R}(\vec x)$. 

We emphasize that this is true for any particular realization of a density peak, regardless of the probability density of the random variables $\delta$, $v^i$ etc. As soon as $v^i$ is proportional to $\partial^i\delta$, diagonalizing $\zeta_{ij}(\vx_\pk)$ will automatically diagonalize $\widetilde v_{kl}(\vx_\pk)$.

As we can see from Eq. (\ref{eq:spinsecondorder}), the spin is also zero at second-order and, presumably, at any higher order since the relation $v^i\propto\partial^i\delta$ holds at any order in perturbation theory as long as $k\ll \HH$. Furthermore, it is also true in the more familiar Newtonian longitudinal gauge where $\delta=4{\cal R}/3$ and the velocity is proportional to $\partial^i {\cal R}$ and therefore to
$\partial^i\delta$.

\section{The PBH spin  after horizon crossing}
\label{sec:throughcrossing}
From horizon crossing onwards, i.e. when the characteristic wavelength of the perturbations becomes smaller than the Hubble radius, the relation between the velocity and the overdensity changes. As a result, the spin can grow briefly owing to the linear tidal torque until the perturbation decouples from the background. Equating the free-fall and sound crossing timescale, the Jeans criterion tells us that gravitational instability occurs for perturbation with comoving wavenumber
\be
\frac{k}{\cal H} < \sqrt{\frac{2\pi}{3}} \frac{(1+\delta_\pk)^{1/2}}{c_{\text{\tiny S}}} \approx \frac{2}{c_{\text{\tiny S}}} ,
\ee
up to a factor of order unity. We have assumed $\delta_\pk \sim 1$ in this estimate.
Since the sound speed satisfies $c_{\text{\tiny S}}\simeq 1/\sqrt{3}$ deep in radiation domination, the perturbations must decouple from the background around horizon crossing in order to form PBHs. Otherwise, radiation pressure would quickly stabilise them. Assuming that turnaround occurs at horizon crossing, we can estimate the amount of angular momentum acquired by the perturbation through linear tidal-torque.

To be consistent with the previous section, we will work in the CMC slicing. Since the reader might be more familiar with the Newtonian longitudinal gauge we first summarise the basic steps to go from the Newtonian longitudinal gauge to the CMC gauge.

\subsection{Relating Newtonian longitudinal gauge and the CMC gauge}
To relate the CMC gauge and the Newtonian slicing and longitudinal gauge we follow the standard reference \cite{bardeen}.
The subset of metric perturbations we will be concerned with are 
\begin{eqnarray}
 g_{00}&=&-a^2(\eta)\left[1+2AQ^{(0)}(x^\mu)\right],\nonumber\\
 g_{0i}&=&-a^2(\eta)BQ_i^{(0)}(x^\mu),\nonumber\\
 g_{ij}&=&a^2(\eta)\left[1+2 H_LQ^{(0)}(x^\mu)\delta_{ij}+2H_TQ_{ij}^{(0)}(x^\mu)\right],
\end{eqnarray}
where $Q^{(0)}(x^\mu)$ is the scalar harmonic and $Q_i^{(0)}(x^\mu)$ and $Q_{ij}^{(0)}(x^\mu)$ are the corresponding vector and tensor. Taking a coordinate transformation of the
form
\begin{eqnarray}
\widetilde{\eta}&=&\eta+T(\eta)Q^{(0)}(x^\mu),\nonumber\\
\widetilde{x}^i&=&x^i+L(\eta)Q^{(0)i}(x^\mu),
\end{eqnarray}
one gets the following transformations 
\begin{eqnarray}
\widetilde{A}&=&A-T'-{\cal H} T,\nonumber\\
\widetilde{B}&=&B+L'+k T,\nonumber\\
\widetilde{H}_L&=&H_L-{\cal H} T-\frac{k}{3}L,\nonumber\\
\widetilde{H}_T&=&H_T + k L.
\end{eqnarray}
We now impose the CMC gauge to be characterised by  uniform-Hubble-constant hypersurfaces, that is 
\be
H_L^{'\CMC}+\frac{k}{3}B^\CMC-{\cal H} A^\CMC=0,
\ee
and we impose this condition going from the Newtonian longitudinal gauge to the CMC gauge. In the Newtonian longitudinal gauge we use the notation
of Ref. \cite{Bartolo:2006fj}, for which $A=\Phi$, $B=H_T=0$ and $H_L=-\Psi=-\Phi$. We omit labels to designate longitudinal gauge quantities for the
purpose of avoiding clutter.
Furthermore, we have assumed zero shear or, equivalently, zero anisotropic stress (which is an excellent approximation at PBH formation
  since the mean free path is much shorter than any relevant scale.)

In a radiation-dominated universe the  corresponding  amplitude of change in the time coordinate is
\be
T=3\left(k^2+6{\cal H}^2\right)^{-1}\left(\Phi'+{\cal H}\Phi\right)=\frac{6\HH^2}{k^2+6\HH^2}\frac{v}{k},
\ee
where the amplitude of the velocity is $v^i=-i(k^i/k)v$.
This fixes the time-slicing.
The corresponding change of the overdensity is then
\be
\label{a1}
\delta_\CMC=\delta+4{\cal H}T=\delta+\frac{12{\cal H}}{k^2+6{\cal H}^2}\left(\Phi'+{\cal H}\Phi\right).
\ee
On superhorizon scales where $\Phi(\vec k,0)=-2{\cal R}/3$ does not depend on time, this gives 
\be
\delta_\CMC(\vec k,0)=\delta(\vec k,0)+2\Phi(\vec k,0).
\ee
Since $\delta_\CMC(\vec k,0)=(2/3)(k/{\cal H})^2{\cal R}$, one correctly reproduces the condition $\delta(\vec k,0)=-2\Phi(\vec k,0)$ expected
in the limit $k\ll {\cal H}$ for the Newtonian longitudinal gauge.
Physically, this reflects the fact that a constant shift in the gravitational potential does not yield any observable effect. In other words,
  PBHs must trace peaks of the familiar (i.e. CMC) density field rather than peaks of the gravitational potential.

While we have fixed the CMC slicing, we have not determined the threading yet. We can do so by imposing $B^\CMC=0$
\be
B^{\CMC}=0+L'+k T=0,
\ee
where we have inserted the zero to remind the reader that $B=0$ in the longitudinal gauge. 
This fixes 
\be
L'=-kT.
\ee
 Given the fact that the velocity transforms as
\be
\widetilde{v}=v+L',
\ee
the gauge-invariant velocity is $(v-H_T'/k)$. This leads to 
\be
v^\CMC-\frac{H_T^{'\CMC}}{k}=v-\frac{H'_T}{k}.
\ee
This means that the amplitude of the velocity  in the CMC gauge is
\be
v^\CMC=\frac{H_T^{'\CMC}}{k}+v+0,
\ee
where we have highlighted the fact that in the Newtonian longitudinal gauge $H_T=0$. Since
\be
\frac{H_T^{'\CMC}}{k}=0+L'=-kT,
\ee
we finally get 
 \cite{bardeen}
\be
\label{a2}
v^\CMC=v-kT=\frac{(k/{\cal H})^2}{6+(k/{\cal H})^2}v.
\ee
Since on superhorizon scales $v(\vec k,0)\simeq -(1/3)(k/{\cal H}){\cal R}$, we get, in the same limit, 
\be
v^\CMC(\vec k,0)\simeq-\frac{1}{18}\left(\frac{k}{\HH}\right)^3{\cal R}=-\frac{1}{12}\left(\frac{k}{\HH}\right)\delta_\CMC(\vec k,0),
\ee 
which nicely reproduces the result mentioned in the previous Section. On subhorizon scales the CMC velocity coincides of course with the Newtonian longitudinal velocity.

\subsection{Subhorizon description  in the CMC gauge}
The expressions (\ref{a1}) and (\ref{a2}) allow us to compute the overdensity and the velocity shear in the CMC gauge once the perturbations re-enter the horizon, in terms of quantities known in the Newtonian longitudinal gauge.

From now on we consider a  perturbation with characteristic comoving scale entering the horizon at $k_\H\gg k_{\rm eq}$, which is the regime relevant to PBH formation deep in radiation domination.
Here, since the mean free path is very short, one can again ignore any anisotropic stress and take the Bardeen potentials to be $\Phi=\Psi$. 
Near horizon crossing  the radiation density perturbation  reads  \cite{Bartolo:2006fj}
\be
\delta(\vk,\eta) \simeq -6\Phi(\vk,0) \cos(k c_{\text{\tiny S}} \eta) + 4 \Phi(\vk,\eta),  
\ee
where 
\be
\Phi(\vk,\eta)= 3\Phi(\vk,0)\, \frac{\sin(kc_{\text{\tiny S}}\eta)-(kc_{\text{\tiny S}}\eta)\cos(kc_{\text{\tiny S}}\eta)}{(kc_{\text{\tiny S}}\eta)^3}.
\ee
Correspondingly, the Fourier modes of the radiation bulk velocity are given by 
\be
v^i(\vk,\eta) = i\frac{9}{2} \frac{k^i}{k}\Phi(\vk,0)\, c_{\text{\tiny S}}\sin(k c_{\text{\tiny S}} \eta) \;.
\ee
 This expression holds as long as $k\gtrsim k_{\rm eq}$. 
Since the comoving Hubble parameter satisfies $\HH = \eta^{-1}$, then horizon crossing occurs at $\eta_\H =  k_\H^{-1}$, with $k_\H\propto M_{\PBH}^{-1/2}$. The multiplicative sine factor in the expression of the radiation bulk velocity becomes
\be
\sin(k c_{\text{\tiny S}} \eta_H) = \sin(k/k_{\text{\tiny S}}) \qquad \mbox{with}\qquad k_{\text{\tiny S}} \equiv \frac{k_\H}{c_{\text{\tiny S}}}.
\ee
For $k=k_\H$, we have $\sin(k_\H/k_{\text{\tiny S}}) \simeq 0.546$.
Furthermore, the (physical) velocity shear at horizon crossing is given by
\begin{align}
  v_{ i}^j(\vk,\eta_\H) &= i k_i v^j(\vk,\eta_\H) = -\frac{9}{2} \frac{k_i k^j}{k} \Phi(\vk,0) c_{\text{\tiny S}}\sin(k/k_{\text{\tiny S}})  ,
\end{align}
while the density perturbation is
\be
\delta(\vk,\eta_\H) \simeq -6\Phi(\vk,0) \cos(k/k_{\text{\tiny S}})+ 4 \Phi(\vk,\eta_\H) .
\ee
Using the expressions (\ref{a1}) and (\ref{a2}) we finally get 
\be
\delta_\CMC(\vec k,\eta_\H)=-6\Phi(\vk,0) \frac{\left[2 \left(3 c_{\text{\tiny S}}^2+1\right)+(k/k_{\text{\tiny S}})^2\right]
   \cos \left(k/k_{\text{\tiny S}}\right)-2 \left(3 c_{\text{\tiny S}}^2+1\right) (k_{\text{\tiny S}}/k) \sin
   \left(k/k_{\text{\tiny S}}\right)}{6c_{\text{\tiny S}}^2 +(k/k_{\text{\tiny S}})^2},
\ee
and
\be
 v_{ \CMC i }^j(\vk,\eta_\H) = -\frac{9}{2} \Phi(\vk,0)\frac{k_i k^j}{k} c_{\text{\tiny S}} \lp\frac{k}{k_{\text{\tiny S}}}\rp^2 \frac{\sin \left(k/k_{\text{\tiny S}}\right)}{6 c_{\text{\tiny S}}^2
   +\lp k/k_{\text{\tiny S}}\rp^2}.
\ee
At this point, it is convenient to introduce a normalised density $\nu$, shear $\widetilde v_{ij}$ and Hessian $\zeta_{ij}$ to retain the analogy with \cite{torques}:
\begin{align}
  \label{eq:normalisedvariables}
  \sigma_{\delta_\CMC}\nu(\vx,\eta_\H) &=
  \frac{V}{(2\pi)^3}\int \d^3k\,\left(\frac{k}{k_{\text{\tiny S}}}\right)^2\,T_\delta(k,\eta_\H)\,\Phi(\vk,0)\,W(k)\, e^{i \vec{k} \cdot \vec{x}}, 
  \nonumber\\
\zeta_{\CMC ij}(\vx,\eta_\H) &=
  -\frac{V}{(2\pi)^3}\int \d^3k\,k_i k_j\,  \delta_\CMC(\vk,\eta_\H)
  \,W(k)\, e^{i \vec{k} \cdot \vec{x}},
  \nonumber \\
 v_{\CMC i}^j(\vx,\eta_\H) &= - k_\H
  \frac{V}{(2\pi)^3}\int \d^3k\,\frac{ k_i k^j}{k^2}
  \,\frac{T_v(k,\eta_\H)}{T_\delta(k,\eta_\H)}\,  \delta_\CMC(\vk,\eta_\H)\,W(k)\, e^{i \vec{k} \cdot \vec{x}}
  \equiv g_v(\eta_\H)\widetilde v_{\CMC i}^j(\vx,\eta_\H).
  \end{align}
Here, $W(k)$ is the Fourier transform of a spherically symmetric window function with characteristic wavenumber $k_\H$, whereas the ``transfer functions'' $T_\delta(k,\eta_\H)$ and $T_v(k,\eta_\H)$ are 
\be
 T_\delta(k,\eta_\H)=-6\left(\frac{k_{\text{\tiny S}}}{k}\right)^2 \frac{\left[2 \left(3 c_{\text{\tiny S}}^2+1\right)+(k/k_{\text{\tiny S}})^2\right]
   \cos \left(k/k_{\text{\tiny S}}\right)-2 \left(3 c_{\text{\tiny S}}^2+1\right) (k_{\text{\tiny S}}/k) \sin
   \left(k/k_{\text{\tiny S}}\right)}{6c_{\text{\tiny S}}^2 +(k/k_{\text{\tiny S}})^2}
\ee
and
\be
T_v(k,\eta_\H) &= \frac{9}{2} \lp \frac{k}{k_{\text{\tiny S}}} \rp \frac{\sin \left(k/k_{\text{\tiny S}}\right)}{6 c_{\text{\tiny S}}^2
   +(k/k_{\text{\tiny S}})^2}.
\ee
Notice also that Eq.~\eqref{eq:normalisedvariables} makes it explicit that in the sub-horizon regime the velocity shear is not proportional to the gradients of the density contrast.
As we can see from Eq. (\ref{eq:normalisedvariables}), the rms variance $\sigma_{\delta_\CMC}$ and $\sigma_{\zeta_\CMC}$ can be constructed from the spectral moments
\be
\sigma_j^2 \equiv \frac{V}{2\pi^{2}}\int \d k\, k^{2+2j} \,\big\lvert \delta_\CMC (\vk,\eta_\H)\big\lvert^2\,W^2(k) .
\ee
This shows that
\be
\sigma_{\delta_\CMC } = \sigma_0, \qquad \sigma_{\zeta_\CMC}=\sigma_2 ,\qquad \sigma_{\times\CMC} = \sigma_1 .
\ee
Finally we define
\be
g_v^2(\eta_\H) = k_\H^2
\frac{V}{2\pi^{2}}\int \d k\, k^2\frac{T^2_v(k,\eta_\H)}{T^2_\delta(k,\eta_\H)}\, \,\,\big\lvert\delta_\CMC (\vk,\eta_\H)\big\lvert^2\,W^2(k) 
\sim \frac{T^2_v(k_\H,\eta_\H)}{T^2_\delta(k_\H,\eta_\H)}
 k_\H^2 \sigma_{\delta_\CMC }^2 (\eta_\H),
\ee
where we approximated the contribution of the transfer functions to a constant, since in all the relevant cases the power spectrum is peaked at $k= k_\H$. The numerical value is found to be $T_v(k_\H,\eta_\H)/T_\delta(k_\H,\eta_\H)\sim 0.5$.

Since $\widetilde v_{\CMC ij}$ does not vanish and is not aligned with the Hessian of $\delta_\CMC$, some angular momentum is generated at linear order. 
With the assumption that turnaround occurs close to horizon crossing, the corresponding reference spin is
\be
S_{\rm ref}(\eta_\H) = \frac{4}{3} a^4(\eta_\H) \overline{\rho}_{\rm rad}(\eta_\H) g_v(\eta_\H)R_*^5 (1-f)^{5/2}. 
\ee
This determines the amplitude of the Kerr parameter at turnaround as we shall see next.

\begin{framed}
{\footnotesize
\noindent 
We conclude this section by some comments on the computation in the Newtonian longitudinal gauge. To calculate the spin acquired through linear tidal-torque, one should carefully handle the superhorizon contributions. While the velocity shear $\partial_i v^j$ is physical, the superhorizon limit $\delta=-2\Phi(\vk,0)$ is irrelevant as it cannot affect the collapse of the perturbation. Therefore, in the expression of the Hessian of the density field
\begin{align}
  \partial_i\partial_j\delta(\vk,\eta_\H)
  &=  2k_i k_j\Phi(\vk,0)  
 +6 k_i k_j\Phi(\vk,0) \left[\cos(k/k_{\text{\tiny S}})- 2\frac{\sin(k/k_{\text{\tiny S}})-(k/k_{\text{\tiny S}})\cos(k/k_{\text{\tiny S}})}{(k/k_{\text{\tiny S}})^3}-\frac{1}{3}\right] ,
\end{align}
using the first term in the right-hand side to define the shape of the ellipsoid would be equivalent to select peaks of the gravitational potential or, equivalently, the curvature. However, one cannot impose any constraint on the value of the gravitational potential at $\vx=\vx_\pk$ because this would not lead to any observable effect. To remedy this problem, one should  subtract the superhorizon contribution to the density in the calculation of the covariance etc. and consider instead
\begin{align}\label{Delta}
 \delta_\CMC (\vk,\eta_\H) &\equiv \delta(\vk,\eta_\H) +2 \Phi(\vk,0) 
  \nn \\
  &= -6\Phi(\vk,0) \left[\cos(k/k_{\text{\tiny S}})- 2\frac{\sin(k/k_{\text{\tiny S}})-(k/k_{\text{\tiny S}})\cos(k/k_{\text{\tiny S}})}{(k/k_{\text{\tiny S}})^3}-\frac{1}{3}\right] .
\end{align}
As a result, the density Hessian that defines the ellipsoidal shape of the perturbation becomes
\begin{align}
	\label{deltaexp}
  \partial_i\partial_j \delta_\CMC(\vk,\eta_\H) &=
  6 k_i k_j\Phi(\vk,0) \left[\cos(k/k_{\text{\tiny S}})- 2\frac{\sin(k/k_{\text{\tiny S}})-(k/k_{\text{\tiny S}})\cos(k/k_{\text{\tiny S}})}{(k/k_{\text{\tiny S}})^3}-\frac{1}{3}\right] \nn \\
 &=  \frac{13}{5} \left(\frac{k}{k_{\text{\tiny S}}}\right)^2 k_i k_j\Phi(\vk,0) + \dots 
\end{align}
which is suppressed by an additional factor of $(k/k_{\text{\tiny S}})^2$. This would ensure that the high density fluctuations which collapse to form PBHs trace the peaks of the radiation density field rather than the peaks of the curvature. 
}
\end{framed}

\section{An estimate of the  PBH dimensionless Kerr parameter at formation time}
From now on, we will work at first-order in perturbation theory and, therefore, we will remove the label ${}^{(1)}$ to avoid cluttering notation. One dimensionless parameter which is of interest to us is the Kerr parameter  
\be
a_{\rm s}=\frac{S}{G_N M^2}=\frac{S_{\rm ref}(\eta_\H)}{G_NM^2}s_{\rm e} \equiv A(\eta_\H)s_{\rm e},
\ee
such that 
\be
A(\eta_\H) =  \frac{4}{3} \frac{a^4(\eta_\H) g_v(\eta_\H) \overline{\rho}_{\rm rad}(\eta_\H)  R_*^5 (1-f)^{5/2}}{G_N M^2}.
\ee
 Substituting  $a^4(\eta_\H) \overline{\rho}_{\rm rad}(\eta_\H) = \overline{\rho}_{\rm rad}(\eta_0)$ gives
\be
A(\eta_\H) = \frac{4}{3} \frac{ \overline{\rho}_{\rm rad}(\eta_0)  g_v(\eta_\H) R_*^5(1-f)^{5/2}}{G_N M^2},
\ee
where $\eta_0$ identifies the present time. This quantity has been evaluated at the time $\eta_\H$ when the relevant modes of the perturbation re-enter the horizon to collapse into a 
PBH. 
In the subhorizon regime, the leading time dependent factor in the velocity definition $g_v(\eta_\H)$ scales like  $g_v (\eta_\H) =  (1/2) {\cal H}(\eta_\H)\sigma_{\delta_\CMC }(\eta_\H)$.
In the radiation phase the known behaviour $\HH(\eta) \propto a^{-1}(\eta)$ leads to the relation
\be
a(\eta_\H) = a(\eta_{\rm eq}) \frac{\HH(\eta_{\rm eq})}{\HH(\eta_\H)} 
= a^{1/2}(\eta_{\rm eq}) \frac{\HH_0}{\HH(\eta_\H)} \sqrt{2 \Omega_\text{\rm dm}},
\ee
where the subscript ${}_{\rm eq}$ denotes the radiation-matter equality, $\Omega_\text{\rm dm}$ represents the present DM abundance, and where we used the fact that in the matter-dominated  era $\HH \sim a^{-1/2}$. We can compute the Hubble rate at crossing time
\be
\label{aH}
\HH(\eta_\H) 
= a^{1/4}(\eta_{\rm eq}) (2 \Omega_\text{\rm dm})^{1/4}
\sqrt{\frac{\HH_0}{2 G_NM}},
\ee
where we used the relation
\be
\HH(\eta_\H) = \frac{a(\eta_\H)}{2 G_N M},\,\,\,\,M\simeq M_\H\equiv \frac{4\pi}{3}
\overline{\rho}_{\rm rad}(\eta_\H)
\lp \frac{a(\eta_\H)}{\HH(\eta_\H)}\rp ^3
\ee
to express the Hubble rate at the horizon crossing as a function of the primordial black hole mass $M$ and Newton coupling constant $G_N$.
We then find 
\be
A(\eta_\H) \sim  \frac{4}{3}\cdot\frac{1}{2}\frac{ \HH(\eta_\H)\sigma_{\delta_\CMC } (\eta_\H)}{ G_N M^2} 
 \lp \Omega_{\rm rad}\frac{3\HH_0^2}{8\pi G_N} \rp 
 R_*^5(1-f)^{5/2} =
  \frac{1}{4 \pi} \frac{\HH_0^2}{  G_N^2M^2} \Omega_{\rm rad} \HH(\eta_\H)\sigma_{\delta_\CMC } (\eta_\H)R_*^5(1-f)^{5/2}. 
\ee
For a power spectrum peaked at the scale $R_*$ one has typically $(1-f)\sim 1/3$  (in the CMC gauge) and  $R_* \sim\sqrt{3} k_\H^{-1} \sim \sqrt{3}\HH^{-1}(\eta_\H)$ \cite{musco}, such that
\be
A(\eta_\H) \sim \frac{1}{4\pi} \frac{\HH_0^2}{G_N^2M^2} \Omega_{\rm rad}\sigma_{\delta_\CMC }(\eta_\H) \HH^{-4}(\eta_\H)
\ee
and substituting Eq. \eqref{aH} in the previous equation gives
\be
A(\eta_\H) \sim  \frac{1}{4 \pi} \frac{\HH_0^2}{G_N^2M^2} \Omega_{\rm rad} a^{-1}(\eta_{\rm eq}) (2 \Omega_\text{\rm dm})^{-1} \sigma_{\delta_\CMC }(\eta_\H) \frac{4G_N^2M^2}{\HH_0^2}.
\ee
An estimate for the Kerr parameter at the time of formation is, thus, provided by the simple relation
\begin{tcolorbox}[colframe=white,arc=0pt]
\vspace{-.15cm}
\be
\label{a}
a_{\rm s} = A(\eta_\H)s_{\rm e} \sim\llp \frac{1}{2\pi} \sigma_{\delta_\CMC }(\eta_\H)  \rrp s_{\rm e}.
\ee
\end{tcolorbox}
\noindent
Notice that $a_{\rm s}$ scales correctly as a first-order quantity.
To get the feeling of the numbers, we take 
$ \sigma_{\delta_\CMC }(\eta_\H) =\delta^{\rm c}_{\CMC}/\nu \sim 0.08$ 
to get 
\be
a_{\rm s} \sim  10^{-2} s_{\rm e},
\ee
were we used the indicative value $\nu = 8$. The details are presented in Sec.~\ref{sec:ps} and may change upon the shape of the power spectrum.
More importantly, the determination of the probability distribution of $a_{\rm s}$ at the time of formation is directly related that of $s_{\rm e}$.
Therefore, our next step is to study in detail the distribution of $s_{\rm e}$. From Eq. (\ref{lll}) one expects $ s_{\rm e}$ to scale like (for some $i\neq j$) 
 \be
 \label{klo}
 s_{\rm e}\simeq 
 \frac{16 \sqrt{2} \pi }{135\sqrt{3 \Lambda} }   \nu^{5/2}\widetilde{v}_{ij}\alpha_i ={\cal O} (1)\cdot\sqrt{1-\gamma^2},
 \ee
 where we have taken into account that the velocity shear scales like $\sqrt{1-\gamma^2}$ (this point will become more clear in the next section).
As a result, we expect $a_{\rm s}$ to be of the order of 
\be
\label{pp}
a_{\rm s} \sim 10^{-2}\sqrt{1-\gamma^2}.
\ee
 This estimate will be confirmed by the investigation of the probability distribution of $a_{\rm s}$ which will indeed be peaked around the value (\ref{pp}). 
 In particular notice that in the limit of very narrow power spectra, for which $\gamma$ tends to unity, the spin vanishes. This has a physical reason: as we will show in the next section, see Eq. (\ref{al}), the velocity shear has a strong tendency to align itself along the principal axis of the mass ellipsoid due to the strong correlation with the inertia tensor. In the limit $\gamma=1$ this alignment is total and   no spin can be generated.

 \section{The  PBH spin and the statistics of local maxima}
 Following \cite{torques}, the starting point for the calculation of the probability of the spin $s_{\rm e}$ is a joint distribution which involves  the sixteen  variables
\be
\label{16var}
\delta, 
\ \ \ \zeta_i = \frac{ \partial \delta}{ \partial x_i}, \ \ \,
 \zeta_{ij} =\frac{\partial^2 \delta}{ \partial x_i \partial x_j}, 
 \ \ \ v_{i}^j = \frac{\partial v^j}{ \partial x^i},
\ee
where the last two matrices have six components, being both symmetric.
We decide to drop the label CMC from all quantities to simplify the subsequent equations.
We can create a vector $V$ of sixteen components with the joint distribution
\be
f(V_i)\d^{16}V_i = \frac{1}{(2\pi)^8|{\bf M}|^{1/2}}e^{-\frac{1}{2}(V_i-\med{V_i}){\bf M}_{ij}^{-1}(V_j-\med{V_j})}\d^{16}V_i,
\ee
where the covariance matrix $\bf M$ is given by (in the following we will use the fact that $\med{V_i} = 0$)
\be\label{covm}
{\bf M}_{ij} = \langle(V_i-\med{V_i})(V_j-\med{V_j})\rangle.
\ee
In the subhorizon regime, the sixteen variables are correlated as follows
\be
\langle {\delta } ^2 \rangle &= \sigma _{\delta } ^2 ,
\\
\langle {\delta } \zeta_{11} \rangle & = -\langle \zeta_{1} \zeta_{1} \rangle =...= -\sigma_{\times}^2/3,
\\
\langle {\delta } \widetilde v_{11} \rangle & =...= - \sigma _{\delta } /3 ,
\\
\langle \zeta_{11}\zeta_{11} \rangle & = 3 \langle \zeta_{11}\zeta_{22} \rangle=
3 \langle \zeta_{12}^2 \rangle=...= \sigma _{\zeta } ^2/5 ,
\\
\langle \zeta_{11} \widetilde v_{11} \rangle & = 3 \langle \zeta_{11} \widetilde v_{22} \rangle=
3 \langle \zeta_{12} \widetilde v_{12} \rangle=...=  \sigma _{\times } ^2 /5 \sigma _{\delta }  ,
\\
\langle  \widetilde v_{11}^2 \rangle & = 3 \langle  \widetilde  v_{11} \widetilde v_{22} \rangle=
3 \langle  \widetilde v_{12}^2 \rangle= ...= 1/5 ,
\ee
where the ellipsis stands for the other components and the remaining correlators are vanishing.
We can rearrange the variables in the dimensionless forms
\be
\nu &= {\delta }/\sigma_{\delta }, 
\\
x &= -(\zeta_{11} + \zeta_{22} + \zeta_{33})/\sigma_{\zeta},\ \  y = -\frac{1}{2}(\zeta_{11} - \zeta_{33})/\sigma_{\zeta},\ \  z = -\frac{1}{2}(\zeta_{11} -2\zeta_{22} + \zeta_{33})/\sigma_{\zeta}, 
\\
v_A &= -(\widetilde v_{11} + \widetilde v_{22} +\widetilde v_{33}),\ \ 
 v_B = -\frac{1}{2}(\widetilde v_{11} - \widetilde v_{33}),\ \  
v_C = -\frac{1}{2}(\widetilde v_{11} -2 \widetilde v_{22} + \widetilde v_{33}), 
\\
w_1 &= \widetilde v_{23},
 \ \ w_2 = \widetilde v_{13} , 
 \ \ w_3 =\widetilde  v_{12}, 
\\
\widetilde{\zeta}_{1} &
= \zeta_{1}/\sigma_{\times}, \ \ 
\widetilde{\zeta}_{2} = \zeta_{2}/\sigma_{\times}, 
\ \ \widetilde{\zeta}_{3} = \zeta_{3}/\sigma_{\times},
\\
\widetilde{\zeta}_{12} &= \zeta_{12}/\sigma_{\zeta}, \ \ \widetilde{\zeta}_{13} = \zeta_{13}/\sigma_{\zeta}, \ \ \widetilde{\zeta}_{23} = \zeta_{23}/\sigma_{\zeta}, 
\ee
which are correlated as follows 
\be
\label{al}
\langle x^2 \rangle &=\langle \nu^2 \rangle = \langle v_A^2 \rangle = 3 \langle \widetilde\zeta  _1 ^2  \rangle 
= 15 \langle \widetilde \zeta _{12}^2 \rangle = 15 \langle w_3 ^2 \rangle =\langle v_A \nu \rangle=...=1,
\\
\langle x \nu \rangle &= \langle x v_A \rangle=5 \langle v_C z \rangle=15 \langle v_B y \rangle = 15 \langle \widetilde \zeta _{12} w_3  \rangle=...=\gamma,
\\
\langle z^2 \rangle &= 3 \langle y^2 \rangle = \langle v_C ^2 \rangle = 3 \langle v_B^2 \rangle = 1/5,
\ee
where $\gamma = \sigma_{\times}^2/\sigma_{\delta } \sigma_{\zeta}$.
Again, the zero correlators are not reported here.
Since $\langle v_A \nu \rangle^2 = \langle v_A^2 \rangle \langle \nu^2 \rangle = 1$, then the variables $v_A$ and $\nu$ are correlated, so we can drop $v_A$ reducing to only fifteen  independent variables. Notice that in the limit $\gamma=1$ the velocity shear is totally aligned with the inertia tensor.

Focusing now on the principal axes of the matrix $-\zeta_{ij}/\sigma_{\zeta}$, we can use its three eigenvalues plus the three Euler angles, describing the orientation of the principal axes, as new variables in the place of the $\zeta_{ij}$. In particular
\be
x = \lambda_1+\lambda_2+\lambda_3, \ \ y = \frac{1}{2}(\lambda_1-\lambda_3), \ \ z = \frac{1}{2}(\lambda_1-2\lambda_2+\lambda_3).
\ee
Being the system independent on these angles, we can integrate over them, leaving twelve independent variables. The corresponding  distribution is
\be
\label{distribution12}
f(\nu, \widetilde{\zeta}_{i}, \lambda_i, v_B, v_C, w_i) = A e^{-Q_2}|(\lambda_1-\lambda_2)(\lambda_2-\lambda_3)(\lambda_1-\lambda_3)|,
\ee
where we have defined, starting from
\be
\Gamma = \frac{1}{1-\gamma^2},
\ee
\be
A = \frac{5^5 3^{11/2}}{2(2\pi)^{11/2}} \Gamma^3
\ee
and
\be
2Q_2 = \Gamma \nu^2 -2\gamma \Gamma x \nu + \Gamma  x^2 + 15 \Gamma y^2 - 30\gamma \Gamma y v_B + 15 \Gamma v_B^2 + 5\Gamma z^2 -10 \gamma \Gamma z v_C \\
+ 5 \Gamma v_C^2 + 15 \Gamma (w_1^2 + w_2^2 + w_3^2) + 3(\widetilde{\zeta}_{1}^2 + \widetilde{\zeta}_{2}^2 + \widetilde{\zeta}_{3}^2).
\ee
In the region around a peak we can make a Taylor expansion of the kind
\be
\zeta_i=\frac{\partial {\delta }}{\partial x_i} = \left(\frac{ \partial^2 {\delta }}{\partial x_i \partial x_j}\right)\bigg|_\pk (x-x_\pk)^j = \zeta_{ij}|_\pk(x-x_\pk)^j,
\ee
such that
\be
\d^3\widetilde{\zeta}_{i} = \left(\frac{\sigma_{\zeta}}{\sigma_{\times}}\right)^3|\lambda_1\lambda_2\lambda_3|\d^3x_i.
\ee
Integrating the distribution of Eq. (\ref{distribution12})  over space eliminates three variables more, the $\widetilde{\zeta}_{i}$,  and  leaves only nine variables. The distribution obtained in such a way describes the comoving number density of peaks in the element $\d \nu \d v_B \d v_C \d^3w_i$. Furthermore, we can integrate over $v_B$ and $v_C$, as these are unconstrained, leaving finally
\be
\label{peakdis}
N_\pk(\nu, \lambda_i, w_i) \d \nu \d^3\lambda_i \d^3w_i = \frac{B}{R_*^3}e^{-Q_4}F(\lambda_i)\d \nu \d^3\lambda_i \d^3w_i,
\ee
where 
\be
F(\lambda_i) = \frac{27}{2}\lambda_1 \lambda_2 \lambda_3 (\lambda_1-\lambda_2)(\lambda_2-\lambda_3)(\lambda_1-\lambda_3),  \ \ B = \frac{5^43^{9/2}}{2^{11/2}\pi^{9/2}} \Gamma^2,
\ee
and 
\be
2Q_4 = \nu^2 + \Gamma(x-x_*)^2 + 15y^2 + 5z^2+15\Gamma w^2,
\ee
with  $x_* = \gamma \nu$ and $w^2 = w_1^2+w_2^2+w_3^2$. This last quantity identifies the squared radius of a polar coordinate system, with angles $\theta = {\rm arccos}(u)$ and $\phi$, for the variables $w_i$, that we will use from now on.

The next step is to find the probability for the spin $s_{\rm e}$. To do so, we can  rewrite Eq. (\ref{peakdis}) in the following way
\be
N_\pk(\nu, \lambda_i, w_i) \d \nu \d^3\lambda_i \d^3w_i = N_\pk(\nu, \lambda_i, w,u,\phi) w^2 \d w \d \phi \d^3\lambda_i \d \nu \frac{\d u}{\d s_{\rm e}} \d s_{\rm e}
\ee
as a function of the previously defined dimensionless spin through Eq. (\ref{lll})
\be
s_{\rm e} = \frac{2^{9/2}\pi \nu^{5/2}w}{5\times 3^{7/2}\gamma^{5/2}\sqrt{\lambda_1 \lambda_2 \lambda_3}}\sqrt{\beta^2 + (\alpha_3^2-\beta^2)u^2},
\ee
where 
\be
\beta^2(\lambda_i,\phi) = \alpha_1^2 \cos^2 \phi + \alpha_2^2 \sin^2 \phi.
\ee
\begin{framed}
{\footnotesize
\noindent 
In the limit  $\gamma$ very close to unity or, equivalently, $\Gamma\ggg 1$ (that is for a monochromatic spectrum) the
distribution (\ref{peakdis}) for  $w$ approaches a product of  Dirac deltas 
\be
N_\pk(\nu, \lambda_i, \vec w) =\frac{2}{3\cdot 15^3}(2\pi)^2\sigma^2_{\zeta}\frac{B}{R_*^3}y(x-2z)\left[(x+z)^2-9y^2\right](y^2-z^2)
e^{-\lp \nu^2+15 y^2+5z^2 \rp/2 }
\delta(x-x_*)\delta^{(3)}(\vec w).
\ee
As a result,  the value $\vec w=0$ is selected and the off-diagonal terms of the velocity shear vanish and therefore the velocity shear is aligned  with the inertia tensor. This implies that the spin parameter of PBH collapsing from a monochromatic feature is  zero. Furthermore, the expression
(\ref{peakdis}) shows us that the  typical value of the velocity shear scales like $1/\sqrt{\Gamma}=\sqrt{1-\gamma^2}$, a scaling we have been using to estimate $s_{\rm e}$ in Eq. (\ref{klo}).
}
\end{framed}
\noindent
Integrating now over $w$, $\phi$ and $\lambda_i$ (the integration over $\phi$ is replaced then by the integration over $\beta$) gives the distribution of peaks of given height $\nu$ and spin $s_{\rm e}$

\begin{figure}[t]
	\includegraphics[width=0.7 \linewidth]{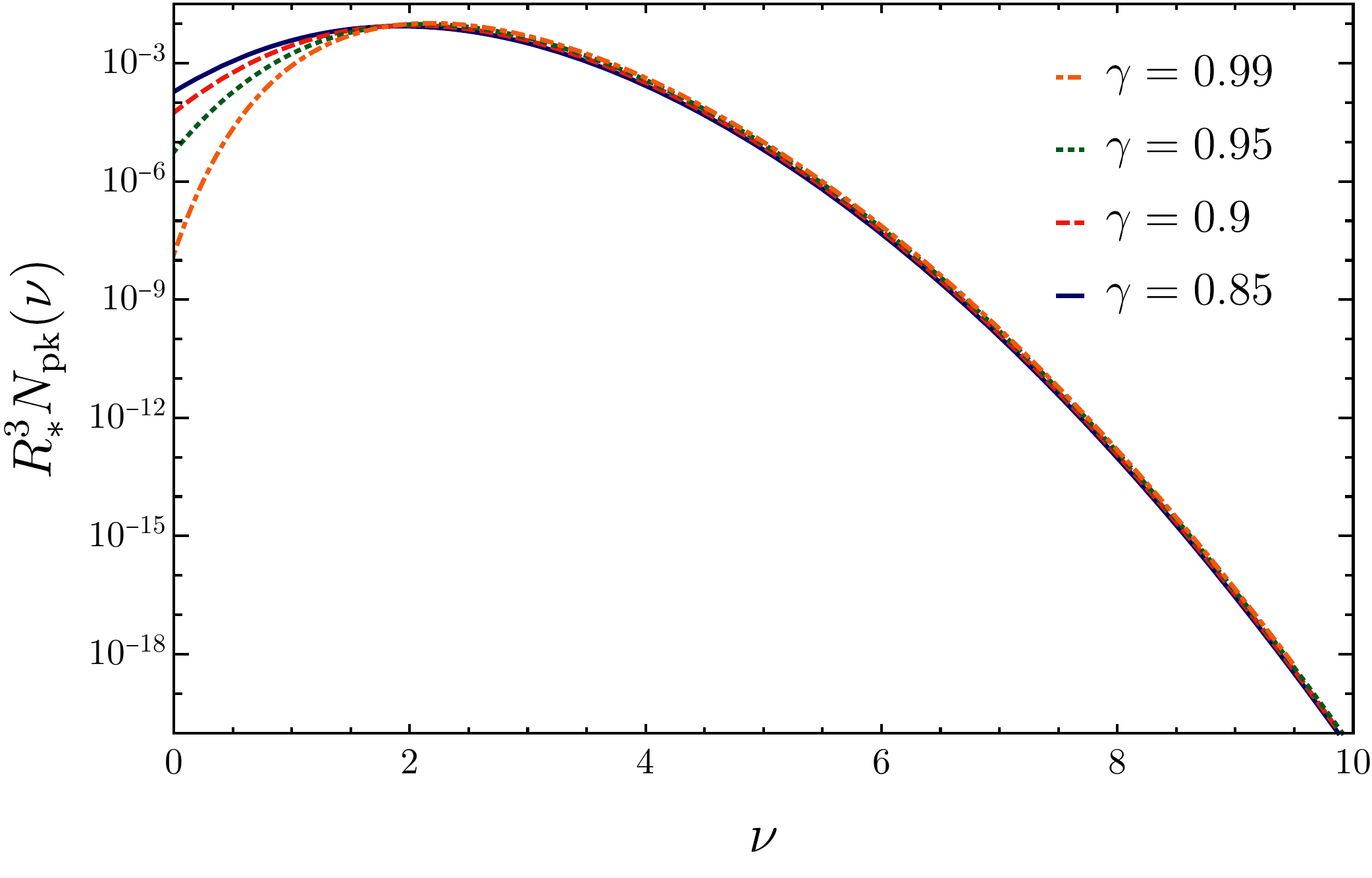}
	\centering
	\caption{The rescaled comoving peak density for several choices of $\gamma$. The curves are almost identical in the relevant region around $\nu \sim 8$. }
	\label{fig:Npeaknu}
\end{figure}

\begin{figure}[t]
	\includegraphics[width=0.49 \linewidth]{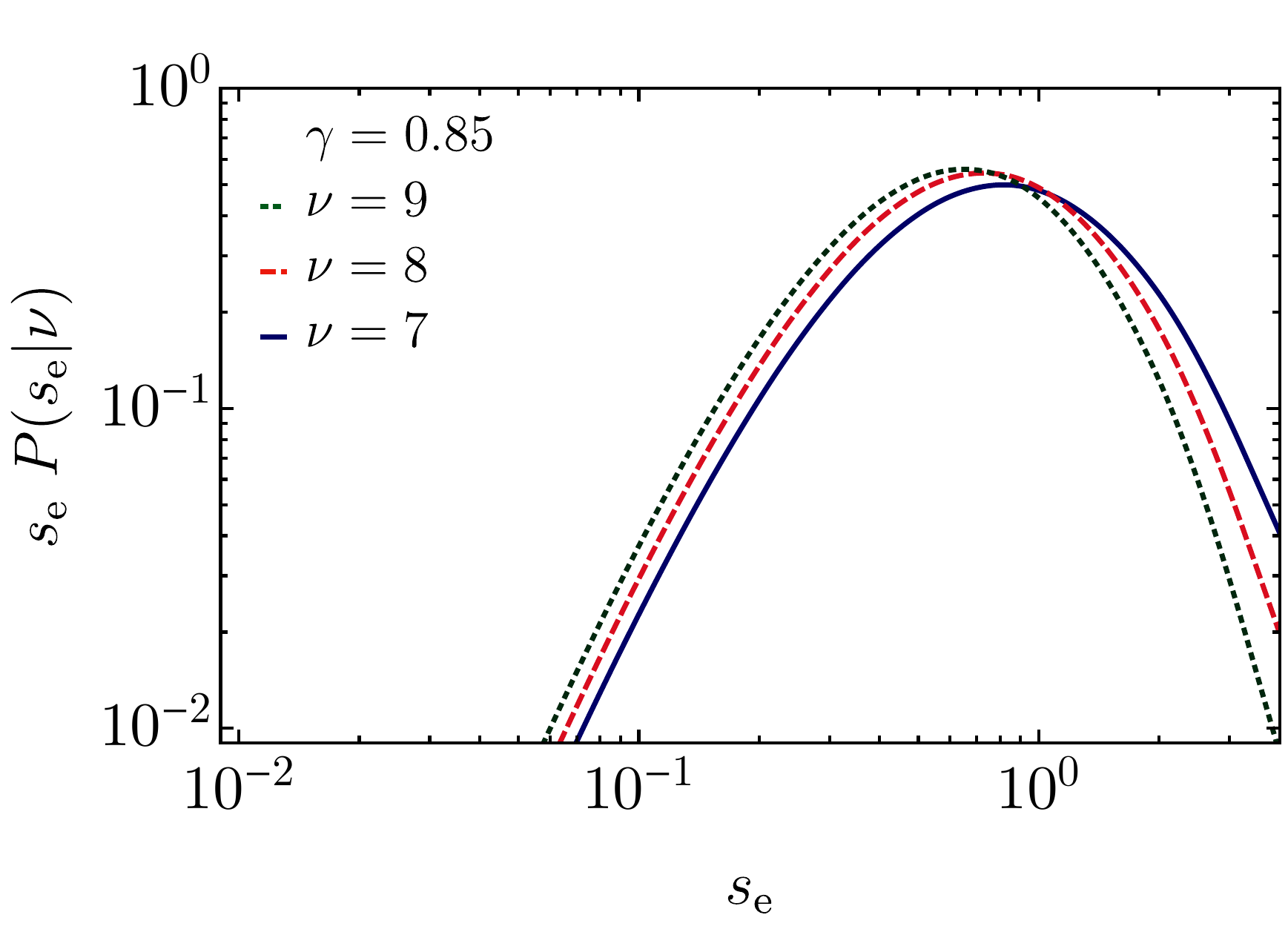}
	\includegraphics[width=0.49 \linewidth]{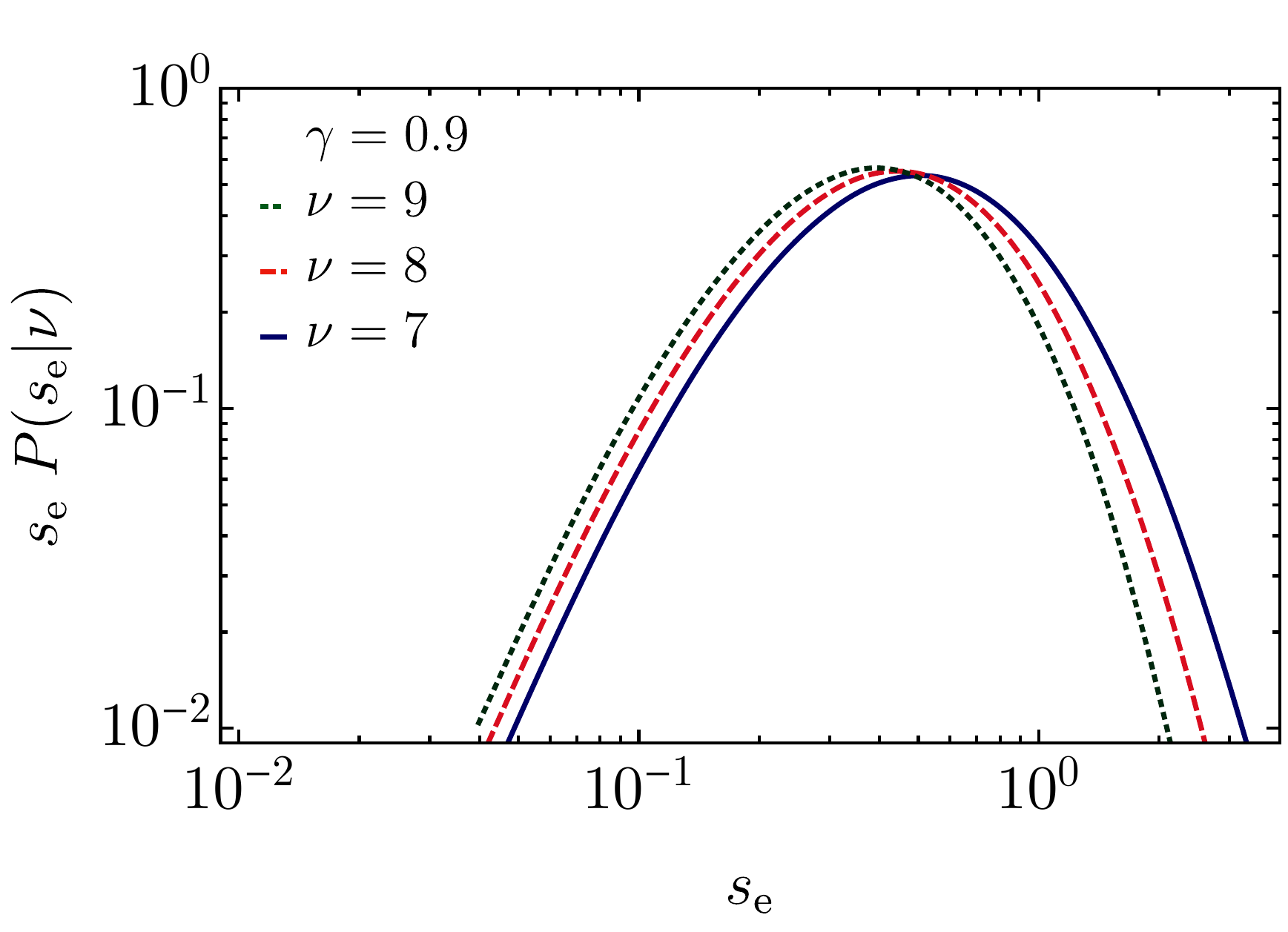}
	\includegraphics[width=0.49 \linewidth]{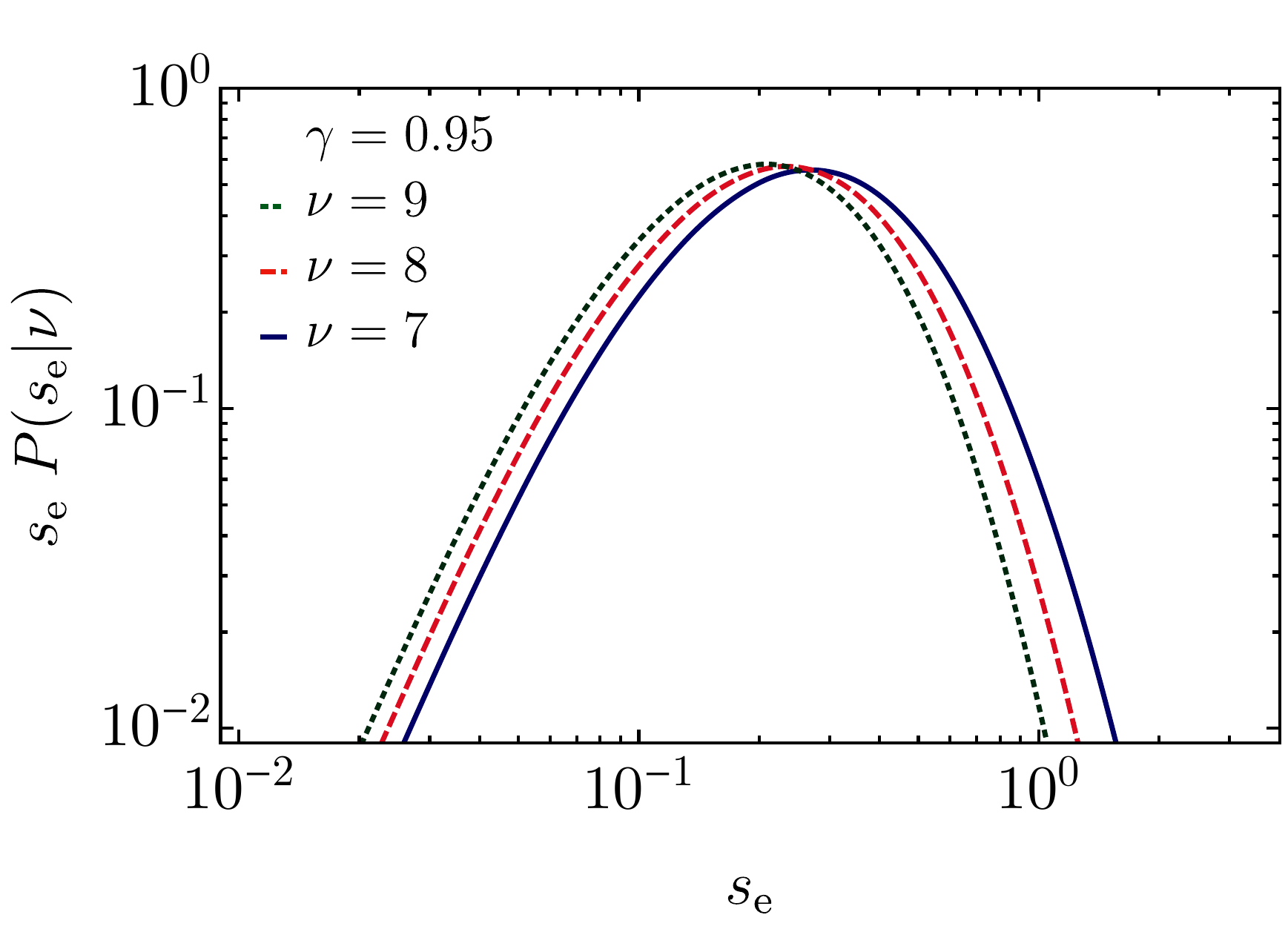}
	\includegraphics[width=0.49 \linewidth]{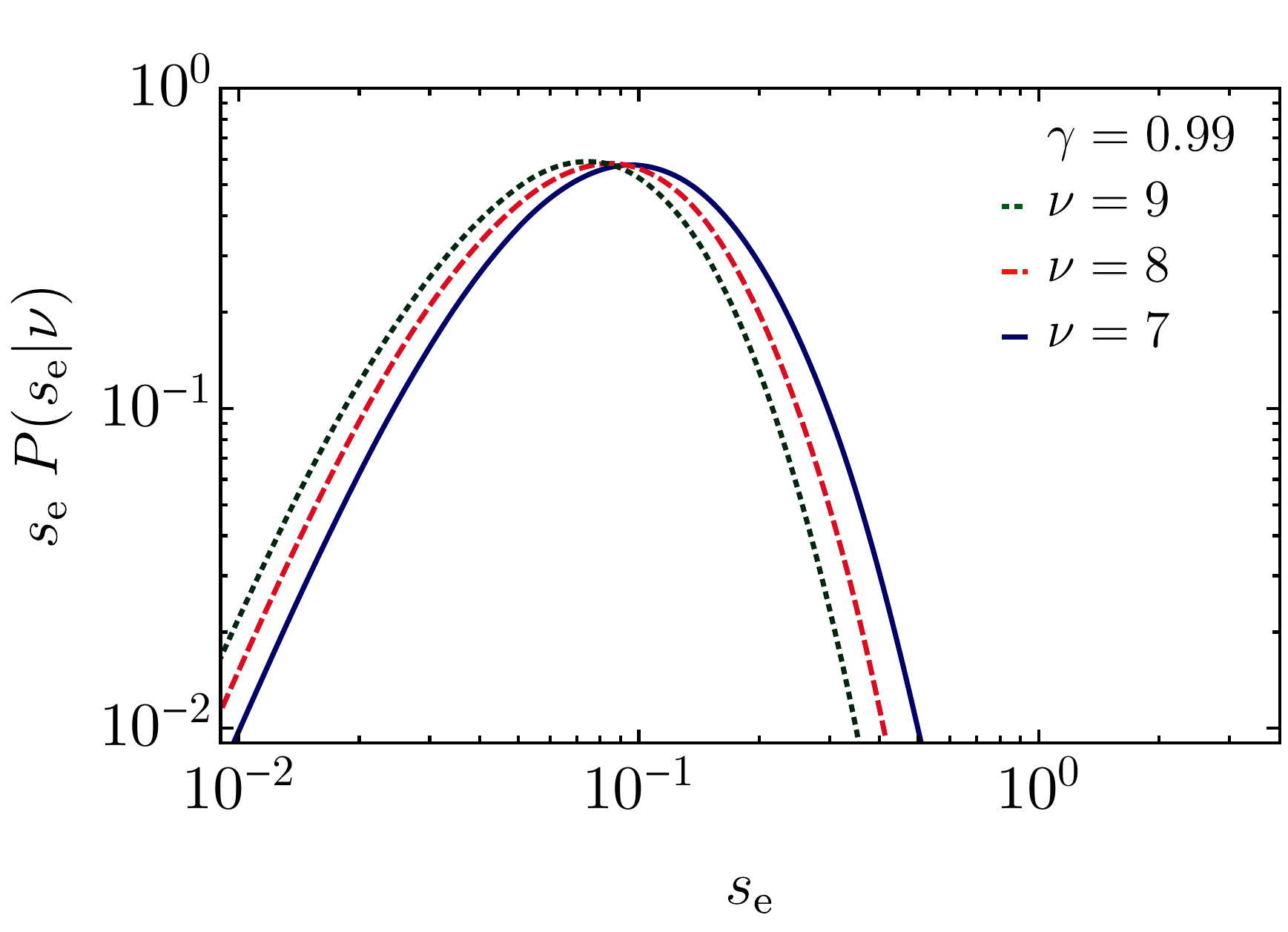}
	\centering
	\caption{Normalised distribution function for $s_{\rm e}$.}
	\label{fig:largenuje}
\end{figure}

\begin{figure}[t]
	\includegraphics[width=0.49 \linewidth]{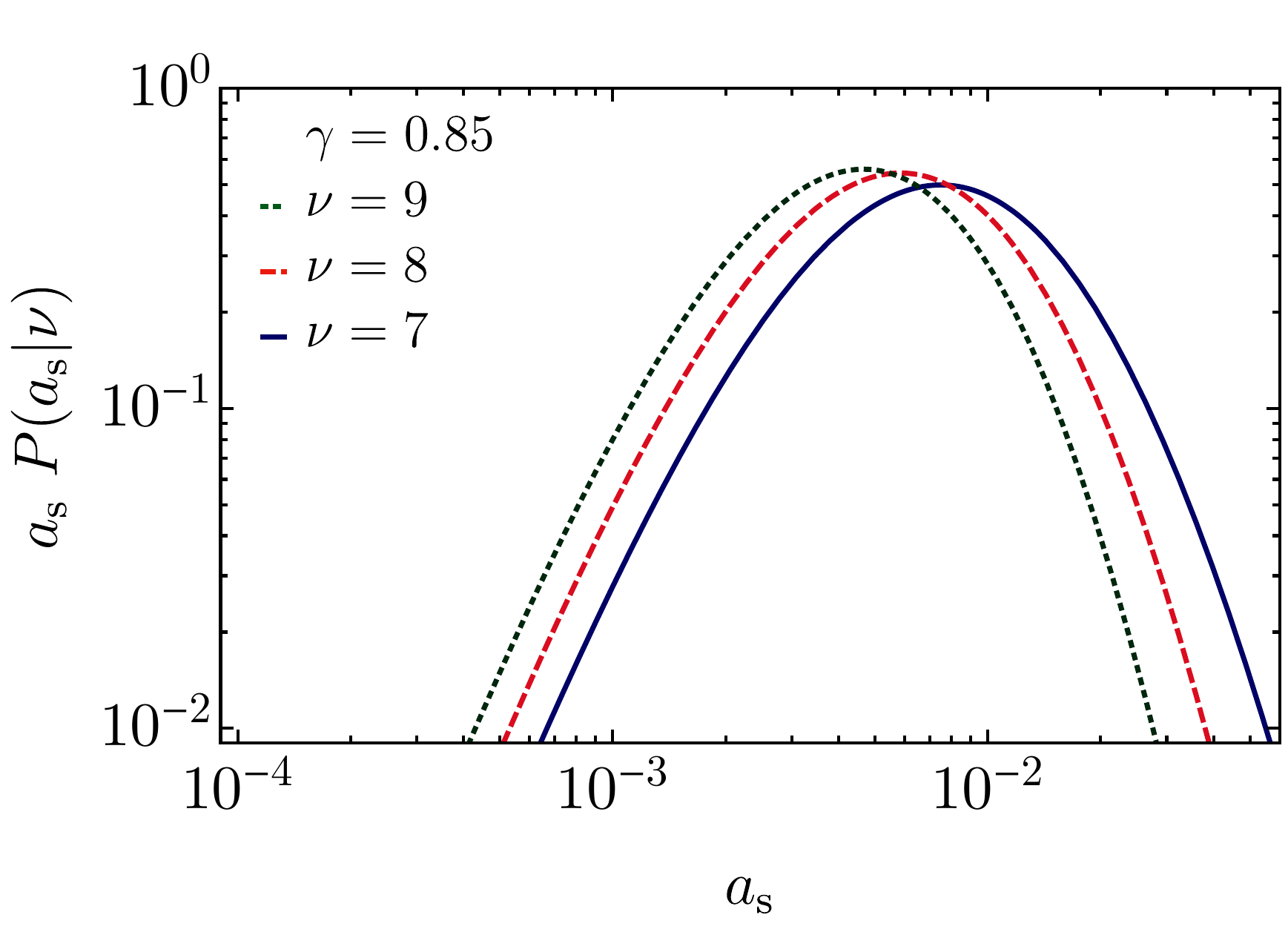}
	\includegraphics[width=0.49 \linewidth]{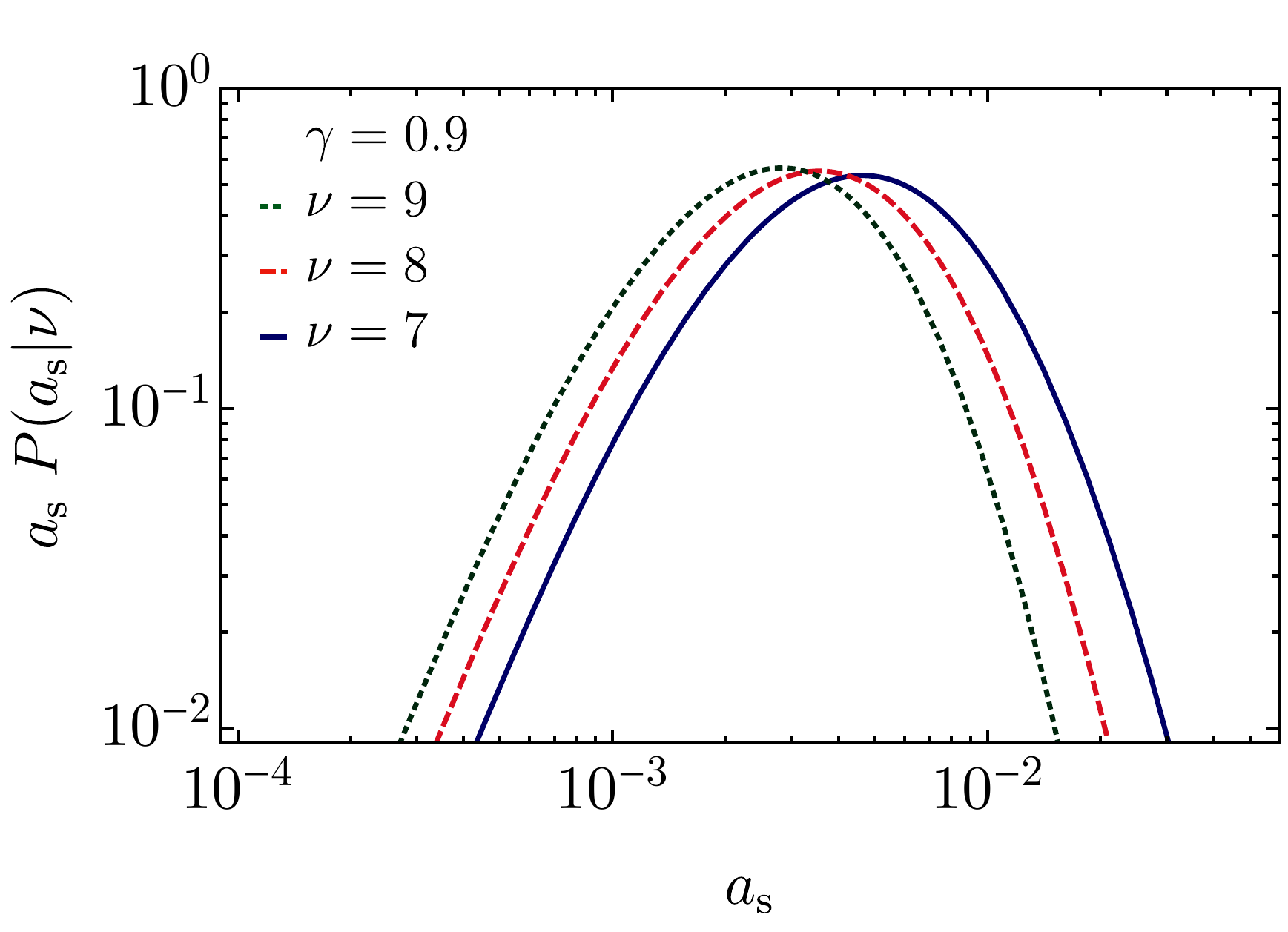}
	\includegraphics[width=0.49 \linewidth]{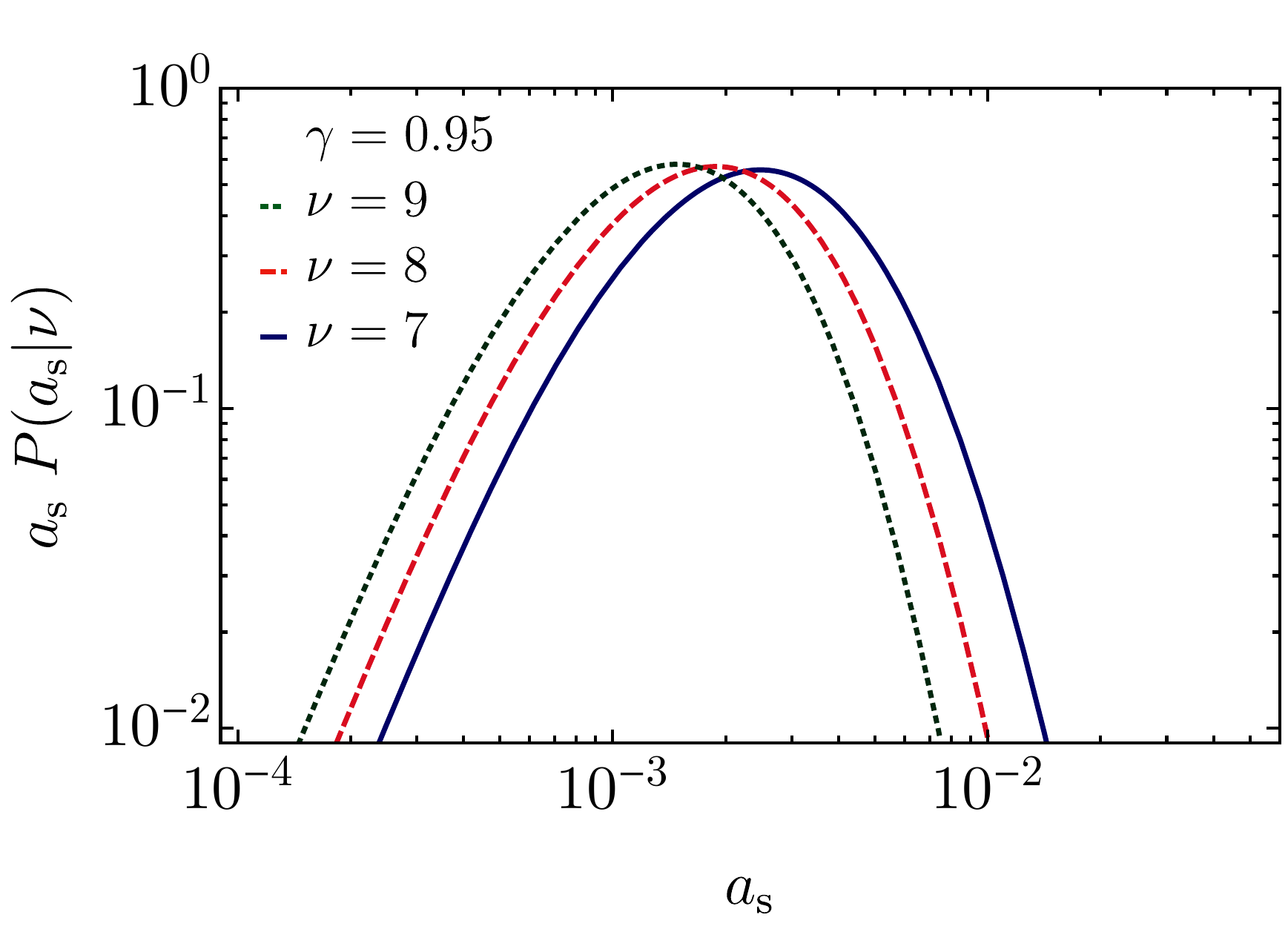}
	\includegraphics[width=0.49 \linewidth]{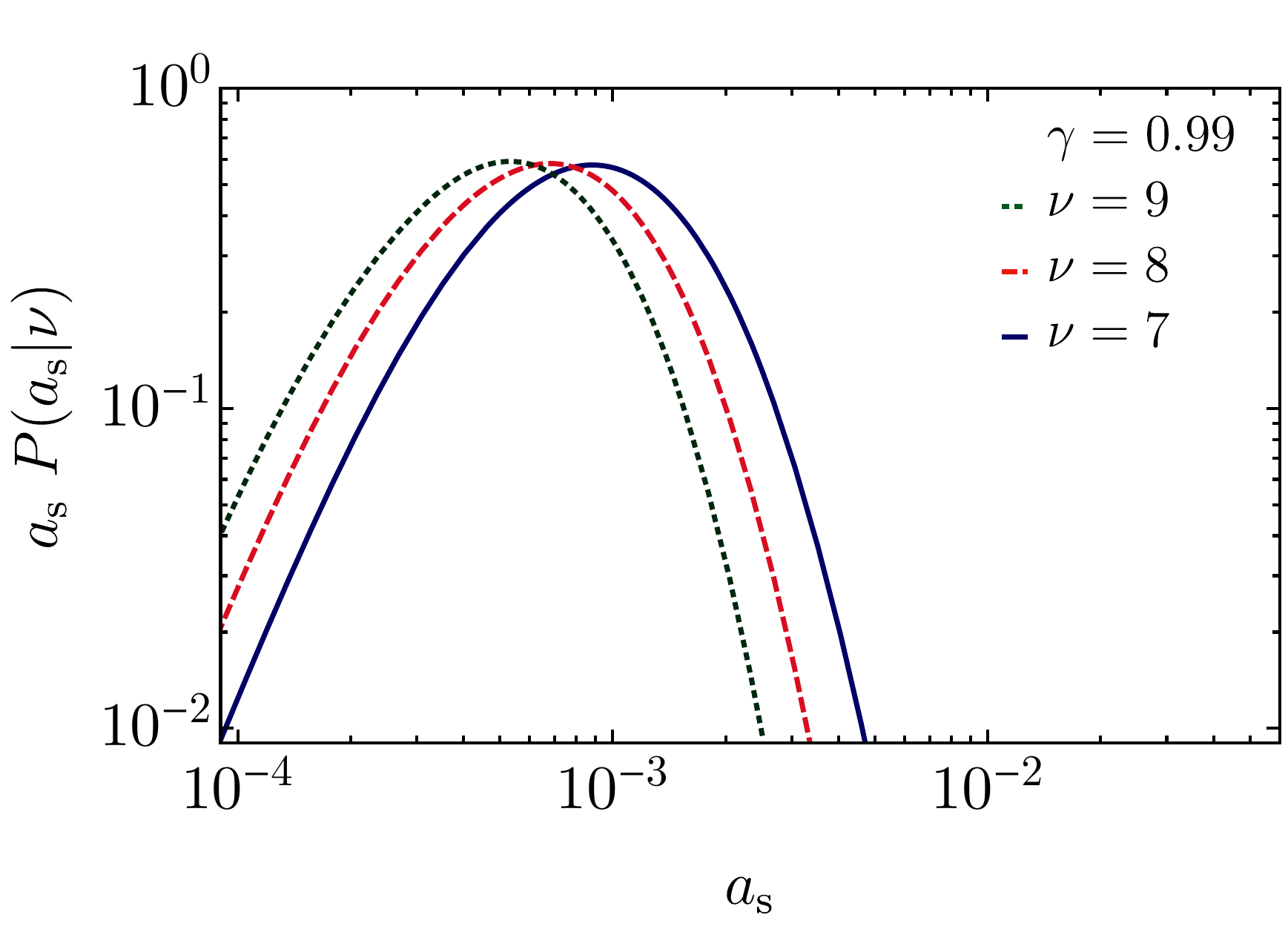}
	\centering
	\caption{Normalised distribution function for $a_{\rm s}$.}
	\label{fig:largenua}
\end{figure}

\begin{tcolorbox}[colframe=white,arc=0pt]
\vspace{-.15cm}
\begin{equation}
\label{Npkvj}
N_\pk(\nu, s_{\rm e}) = \frac{4Cs_{\rm e}}{R_*^3 \nu^5}\int_{0}^{\infty} \d \lambda_1 \int_{0}^{\lambda_1} \d \lambda_2 \int_{0}^{\lambda_2} \d \lambda_3 \int_{\alpha_1}^{\alpha_2} \d \beta \frac{e^{-Q_5}F(\lambda_i)\Lambda T(\alpha_3,\beta, s_{\rm e}, \nu)}{\sqrt{|(\alpha_1^2-\beta^2)(\alpha_2^2-\beta^2)(\alpha_3^2-\beta^2)|}},
\end{equation}
\end{tcolorbox}
\noindent
where 
\be
\Lambda = \lambda_1 \lambda_2 \lambda_3, \ \ C = \frac{3^{11}5^{11/2}\gamma^5\Gamma^{3/2}}{2^{13}\pi^{13/2}}, \ \ 2Q_5 = \nu^2 + \Gamma(x-x_*)^2 + 15y^2 + 5z^2,
\ee
and
\be
T= \Theta(\alpha_3^2-\beta^2)
e^{\frac{-15\Gamma w_3^2}{2}}D(X) + 
\Theta(\beta^2-\alpha_3^2)
\frac{\sqrt{\pi}}{2}
e^{\frac{-15\Gamma w_{\beta}^2}{2}}{\rm erf}(X).
\ee
Here the argument of the Dawson's integral $D(X)$ is given by
\be
X = \sqrt{\frac{15}{2}\Gamma |w_{\beta}^2-w_3^2|}, \ {\rm with} \ w_3=\frac{\sqrt{\Lambda}s_{\rm e}}{K \nu^{5/2}\alpha_3}, \ \ w_{\beta}=\frac{\sqrt{\Lambda}s_{\rm e}}{K \nu^{5/2}\beta}, \ \ K = \frac{2^{9/2}\pi}{5 \times 3^{7/2}\gamma^{5/2}}.
\ee
The conditional differential probability for $s_{\rm e}$, given that the peak has a height $\nu$, is the given by \cite{torques}
\be
\label{prob}
P(s_{\rm e}|\nu) \d s_{\rm e} = \frac{N_\pk(\nu,s_{\rm e})}{N_\pk(\nu)}\d s_{\rm e}, 
\ee
where the comoving differential peak density is expressed as
\be
\label{Npeakn}
N_\pk(\nu) \d \nu = \frac{1}{(2\pi)^2R_*^3}e^{-\frac{\nu^2}{2}}G(\gamma,x_*)\d \nu
\ee
in terms of the function 
\be
G(\gamma,x_*) = \int_{0}^{\infty}\d x f(x) \sqrt{\frac{\Gamma}{2\pi}}e^{-\frac{\Gamma}{2}(x-x_*)^2}.
\ee
The function $f(x)$ that appears in the previous formula is provided by the expression
\be
f(x) = \frac{(x^3-3x)}{2}\bigg[{\rm erf}\left(x\sqrt{\frac{5}{2}}\right) + {\rm erf}\left(\frac{x}{2}\sqrt{\frac{5}{2}}\right)\bigg] + \sqrt{\frac{2}{5\pi}}\bigg[\left(\frac{31x^2}{4} + \frac{8}{5}\right)e^{-\frac{5x^2}{8}}+ \left(\frac{x^2}{2}-\frac{8}{5}\right)e^{-\frac{5x^2}{2}}\bigg].
\ee
Fig.~\ref{fig:Npeaknu} shows the behaviour of the comoving peak density of Eq. \eqref{Npeakn} for several choices of $\gamma$. Fig.~\ref{fig:largenuje} shows the behaviour of the distribution $P(s_{\rm e}|\nu)$ for relevant choices of $\nu$. 
Finally, we can use the relation shown in Eq. \eqref{a} to find the distribution for the Kerr parameter $a_{\rm s}$, plotted in
Fig.~\ref{fig:largenua}. 

We note that the conditional probability distribution for the spin of a material in the vicinity of peaks of different heights shows  a systematic shift towards smaller values for higher peaks if the parameter $\gamma$ is close to unity. 
Furthermore, for a given value of the height peak $\nu$, higher values of $\gamma$ provide slightly smaller  values of spins, and one can check that the scaling with $\gamma$  is indeed like $\sqrt{1-\gamma^2}$. 
\noindent
\noindent

\subsection{The analytical approximation in the  limit of high peaks}
\begin{figure}[t]
	\includegraphics[width=0.7 \linewidth]{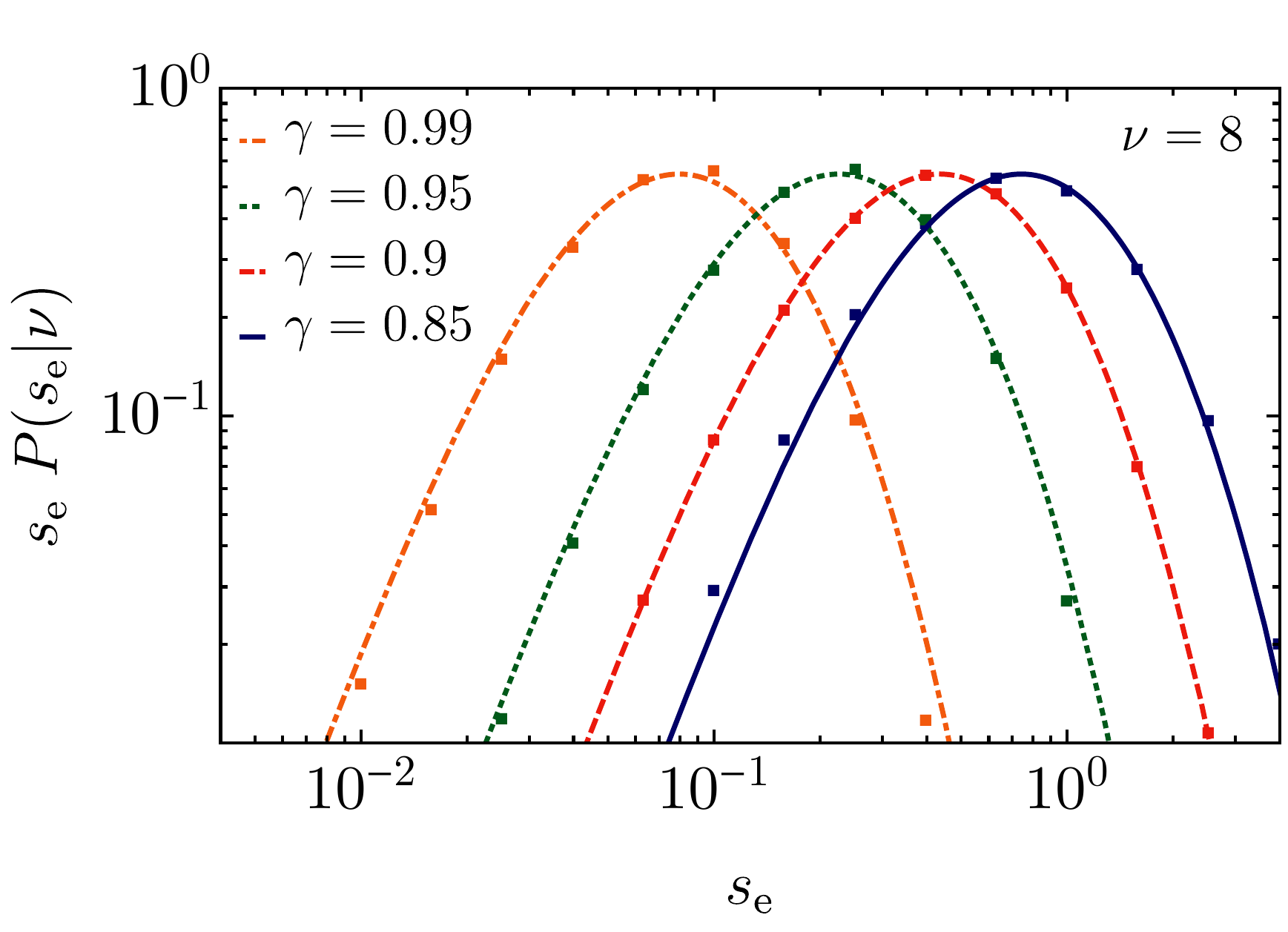}
	\centering
	\caption{Comparison between the numerical result (dots) and the fitted expression (lines) for the probability distribution $P(s_{\rm e}|\nu)$ in the large $\nu$ limit.}
	\label{fig:fit}
\end{figure}

The formation of  PBHs requires high thresholds and it is therefore interesting to investigate analytically the large $\nu$ limit of the probability $P(s_{\rm e}|\nu)$. To do so, we define a normalised dimensionless spin $h$ as\footnote{We note that Eq. (C.7) of Ref. \cite{torques} has the incorrect scaling with $\Gamma$. The factor  $\Gamma^{1/2}$ should be
$\Gamma^{-1/2}$.}
\be
s_{\rm e} \equiv \frac{2^{9/2} \pi}{5 \gamma^6 \nu } \frac{h}{\Gamma^{1/2}}=  \frac{2^{9/2} \pi}{5 \gamma^6 \nu } \sqrt{1-\gamma^2}h. 
\ee
The scaling with $\sqrt{1-\gamma^2}$ is dictated by the scaling of the velocity shear, see Eq. (\ref{klo}). 
The distribution of the parameter $h$ can be analytically approximated as
\be
P(h) \d h=  \exp \lp -2.37 - 4.12 \ {\ln h} - 1.53 \ {\ln^2 h} - 0.13 \ {\ln^3 h} \rp \d h. 
\ee
Such a distribution is the result of a best-fit that was performed for the values $\gamma = 0.9$ and $\nu = 8$. We checked numerically that it holds for the relevant parameter space related to the physics of PBH formation. In Fig. \ref{fig:fit} we compared the numerical result and analytical approximation for the probability distribution $P(s_{\rm e}|\nu)$.

The analytical expression for the distribution of the Kerr parameter $a_{\rm s}$ is found to be in the large $\nu$ limit 
\noindent
\begin{tcolorbox}[colframe=white,arc=0pt]
\vspace{-.4cm}
\label{adistribution}
\begin{eqnarray}
P(a_{\rm s}|\nu)\d a_{\rm s} &=& \lp \frac{5 \gamma^6 \nu }{2^{9/2} \pi} \frac{\Gamma^{1/2}}{A(\eta_\H)} \rp \exp \llp -2.37 - 4.12 \ {\ln \lp \frac{5 \gamma^6 \nu }{2^{9/2} \pi} \frac{\Gamma^{1/2}}{A(\eta_\H)} a_{\rm s} \rp } \right. 
 \nonumber \\
&
- & \left.
 1.53 \ {\ln^2 \lp \frac{5 \gamma^6 \nu }{2^{9/2} \pi} \frac{\Gamma^{1/2}}{A(\eta_\H)} a_{\rm s} \rp} 
 - 0.13 \ {\ln^3 \lp \frac{5 \gamma^6 \nu }{2^{9/2} \pi} \frac{\Gamma^{1/2}}{A(\eta_\H)} a_{\rm s} \rp} \rrp \d a_{\rm s},
\end{eqnarray}
\end{tcolorbox}
\noindent
where $A(\eta_\H)$ is defined in Eq. \eqref{a}.
This is the main result of our paper.
\noindent

\section{The PBH spin and the shape of the power spectrum}\label{sec:ps}
\begin{figure}[t]
	\includegraphics[width=0.6 \linewidth]{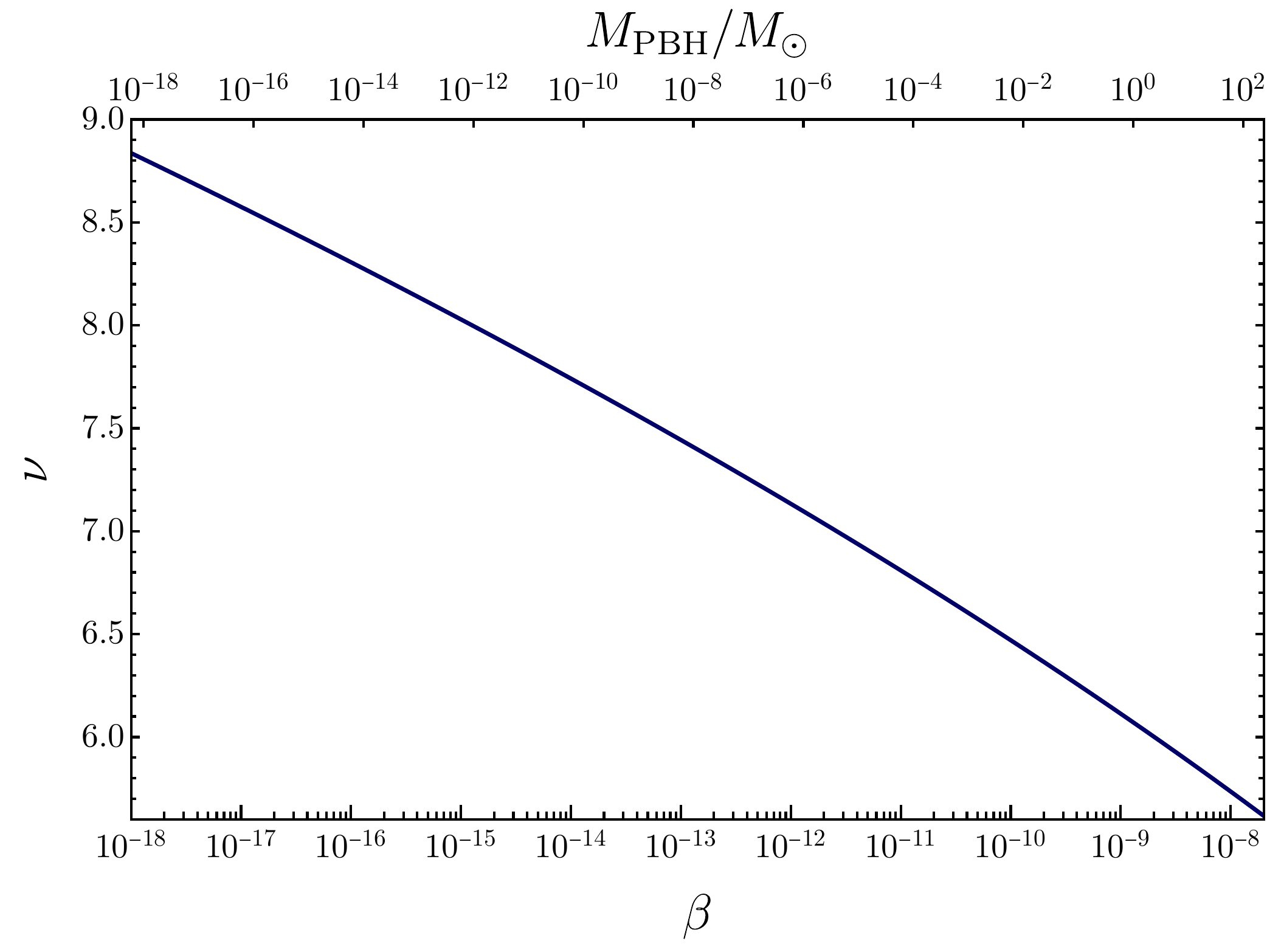}
	\centering
	\caption{The parameter $\nu$ as a function of $\beta$.}
	\label{nubeta}
\end{figure}

In the previous section we have seen that the  spin distribution is  characterised only by two parameters, namely $\nu$ and $\gamma$. The parameter $\nu$ describes the height of the peaks in terms of the variance and  is determined by the required abundance of PBHs. In particular, assuming a Gaussian statistics (for its non-Gaussian extension,  see Ref. \cite{ng}), the PBH mass fraction can be expressed as \cite{sasaki}
\be
\beta = 
\frac{\rho_{\rm PBH}}{\rho_ {\rm rad}}\bigg |_{\rm form}
=
\frac{1}{\sqrt{2 \pi }\nu} \exp\lp- \frac{\nu ^2}{2}\rp 
,
\ee
where $\nu$ is defined as $\nu= \delta^{\rm c}_{\CMC} /\sigma_{\delta_\CMC}$, which is gauge independent.
The approximate relation between $\beta$ and $\nu$ is shown in Fig.~\ref{nubeta}.
\noindent
Assuming a monochromatic spectrum of masses and imposing the PBH to be the dark matter, one may deduce the relation between the mass fraction and the PBH mass to be
\be
\beta\gsim 1.3 \times 10^{-9} \lp \frac{M}{M_\odot}\rp ^{1/2}.
\ee
For models forming PBH in the physically relevant mass range, the parameter $\nu$ lies in the range 
\be
6\lsim \nu \lsim 9.
\ee
 The power spectra  of the density perturbation are  directly calculable in terms of the power spectra of the comoving curvature perturbation $\zeta$ as
\be
{\cal P}_{\delta_\CMC}(k) = \frac{4}{9} \lp \frac{k}{\HH}\rp ^4 {\cal P}_{\zeta,{\rm tot}}(k).
\ee
The quantity  ${\cal P}_{\zeta,{\rm tot}}(k)$ is the sum of two pieces,  the smooth power spectrum giving the correct amplitudes for perturbations at the CMB scales and the term responsible for the formation of PBH at small scales, that is 
\be
{\cal P}_{\zeta,{\rm tot}}(k) = A_s \lp \frac{k}{k_{\rm p}} \rp^{n_\zeta-1} + {\cal P}_\zeta(k),
\ee
where $k_{\rm p}$ denotes a pivot scale, $A_s\sim 2\cdot   10^{-9}$ the corresponding amplitude and $n_\zeta$ the spectral index of scalar perturbations.
To compute the momenta of these distributions, we are going to use the volume normalised Gaussian window function of the form $W(k) = \exp [- (k R_\H)^2 /2]$ smoothing out perturbations on scales different from the characteristic scale $R\H$ corresponding to the cosmological horizon at formation.

These momenta enter in the computation of the parameter $\gamma=\sigma_{\times\CMC}^2/\sigma_{\delta_\CMC } \sigma _{\zeta\CMC}$, which contains all the relevant information on the shape of the power spectrum of the density perturbations. 
In the following subsections we consider few examples of typical power spectra  and compute their characteristic parameter $\gamma$.
\subsection{Log-normal  power spectrum}
\begin{figure}[t]
	\includegraphics[width=0.47 \linewidth]{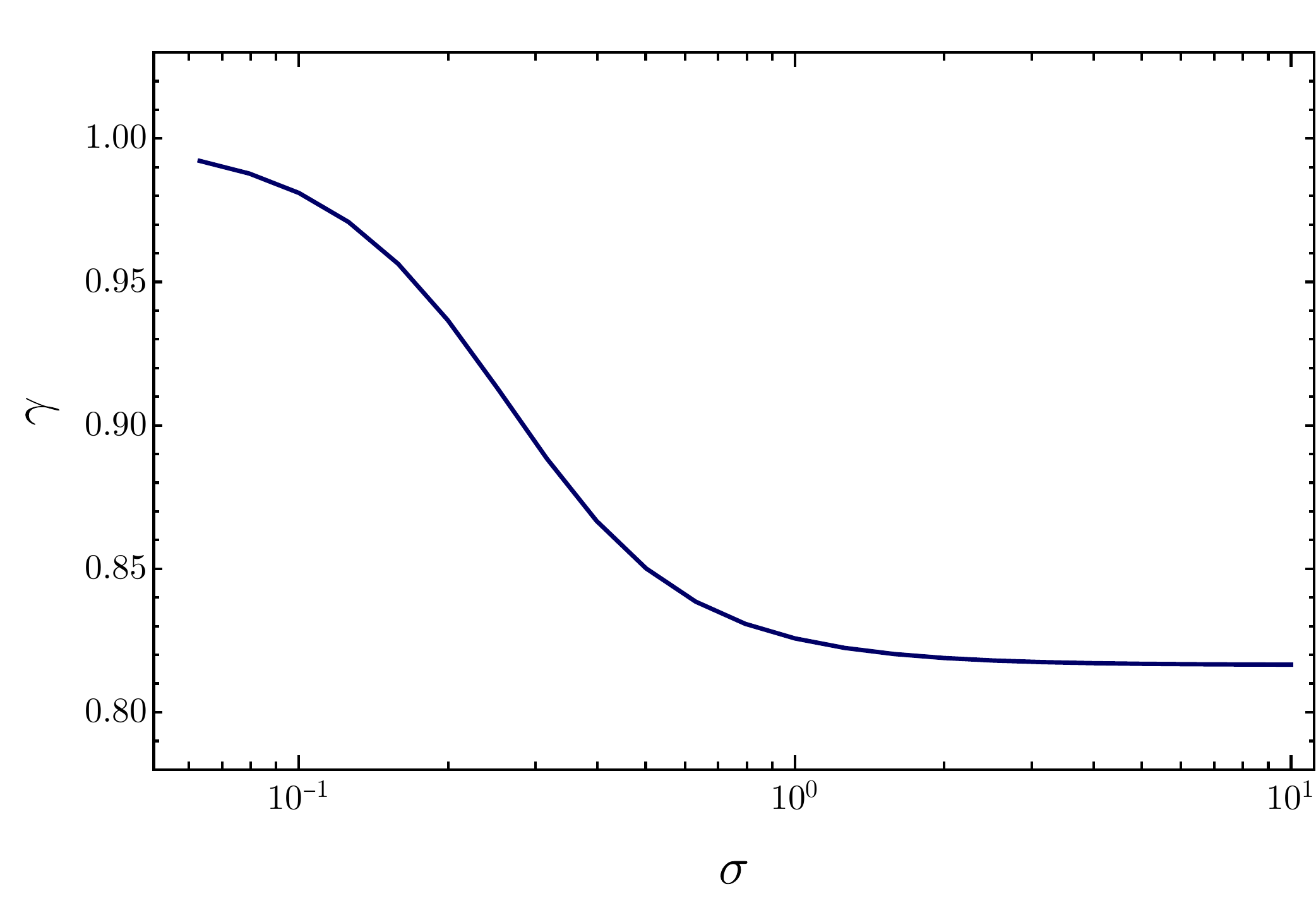}
	\hspace{0.1cm}
	\includegraphics[width=0.47 \linewidth]{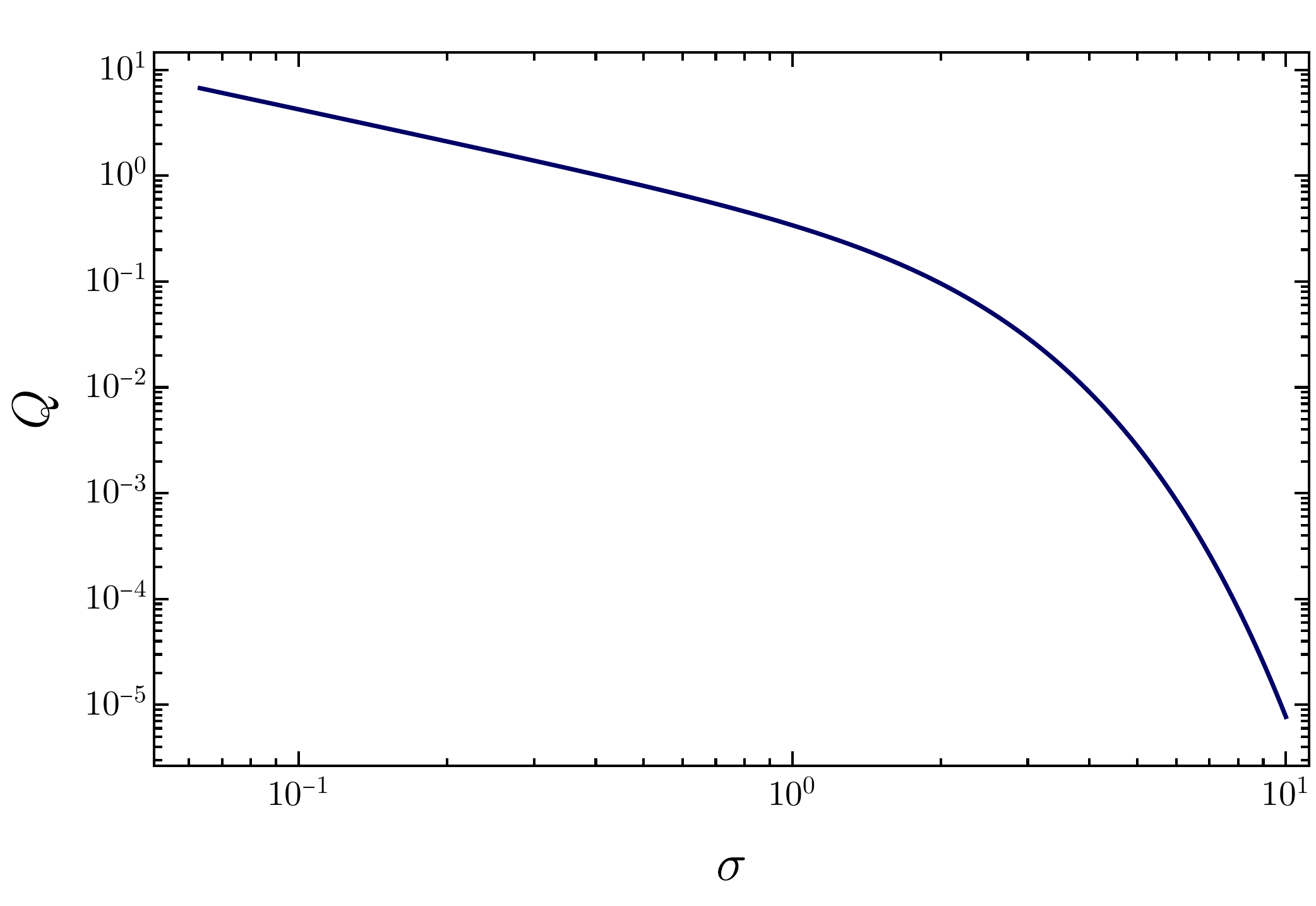}
	\centering
	\caption{Parameter $\gamma$ and $Q$ depending on $\sigma$.}
	\label{figGamma}
\end{figure}

We start by considering  a log-normal power spectrum of the form
\be
{\cal P}_\zeta(k)=A\exp \bigg[-\frac{1}{2\sigma^2}{\log}{\left(\frac{k}{k_*}\right)^2}\bigg].
\ee
In general, the factor $\gamma$ depends only on the parameter $\sigma$ as shown in Fig.~\ref{figGamma}.
\noindent
\hspace{-0.15cm}
We can define the parameter $Q = k_*/\Delta k$ to indicate how peaked is the power spectrum, where $\Delta k$ stands for the full width at half of the maximum value (FWHM).
The value $Q$ for various shapes  is plotted in Fig.~\ref{figGamma}.
For narrow power spectra, the parameter $\gamma$ approaches unity, while for wider profiles it quickly reaches the asymptotic value $\gamma \sim 0.82$.
For example, a choice of parameters consistent with the totality of dark matter  composed by PBH  of mass $M= 10^{-12} M_{\odot}$ would be  $A=0.066$, $\sigma =0.5$ and $k_*= 3 \cdot 10^{12} \,{\rm Mpc}^{-1}$ \cite{noitesting}.

In the limit of small enough $\sigma$, the log-normal power spectrum can also serve as a good approximation of a Dirac power spectrum, usually used in the literature. For example, let us analyse the case of $\sigma = 0.042$ corresponding to a highly peaked power spectrum.
From the explicit expression of the power spectrum we can compute its momenta 
\be
\sigma_{\delta_\CMC }^2 &=  \int \d k {\cal P}_{\delta_\CMC}(k) W^2(k) \frac{1}{k} = 0.0174\,  A^2, \\
\sigma_{\times\CMC}^2 &=   \int \d k {\cal P}_{\delta_\CMC}(k) W^2(k) k =  0.0176 \, k_*^2 A^2, \\
\sigma_{\zeta_\CMC}^2 &=   \int \d k {\cal P}_{\delta_\CMC}(k) W^2(k) k^3=  0.0179 \, k_*^4A^2
\ee
and 
\be
\gamma = \frac{\sigma_{\times\CMC}^2}{\sigma_{\delta_\CMC } \sigma_{\zeta_\CMC}}=0.996  \qquad \text{and} \qquad \Gamma =125.
\ee
Being independent on $A$, our choice of $\sigma$ determines the value $Q = 10$.

\subsection{Nearly flat power spectrum}
For a power spectrum of the form \cite{Byrnes}
\be
{\cal P}_\zeta(k)=A_f \lp \frac{k}{k_{\rm min}}\rp ^{n_{\rm p} -1}  \Theta (k-k_{\rm min}),
\ee
we may fix the parameters such that    PBHs with masses $\sim M_\odot$ form the dark matter. This gives  $n_{\rm p} \sim 0.96$, $A_f= 0.0308$ and $k_{\rm min}= 10^3 {\rm Mpc}^{-1}$. 
For such a case we find
$\gamma \simeq 0.88$.

\subsection{Broken power law power spectrum}
One can also consider the case of a power spectrum of the form \cite{bellido}
\be
{\cal P}_\zeta(k)=A_b \llp \lp \frac{k}{k_{\rm p}}\rp ^{m}  \Theta (k_{\rm p}-k)
+
\lp \frac{k}{k_{\rm p}}\rp ^{-n}  \Theta (k-k_{\rm p}) \rrp,
\ee
where an example of a parameter set giving the PBHs to form the dark matter  is $m=3$, $n=0.5$,  $A_b=0.0413$ and $k_{\rm p}= 2 \cdot 10^6\, {\rm Mpc^{-1}}$, corresponding to a population  of PBH peaked at $M\sim  M_{\odot}$. This set gives $\gamma= 0.85$.

\section{The impact of the spin onto the PBH abundance and the spin distribution function }
Once we have calculated the spin distribution of PBH at formation time, we may study the impact of the spin on the abundance of PBHs. This exercise has been already provided in
Ref. \cite{Jap} where, however, the spin distribution has been assumed to be flat. In order to compare with their results, we are going to assume the same parameter dependence
\be
M = C_{\text{\tiny $M$}} |\delta-\delta_{\rm c}(q)|^{\gamma_{\text{\tiny $M$}}}, \ \ S = c_{\text{\tiny S}} |\delta-\delta_{\rm c}(q)|^{\gamma_S}q, \ \ \delta_{\rm c}(q) = \delta_{0{\rm c}} + Kq^2,
\ee
where $\gamma_{\text{\tiny $M$}} = 0.3558$ \cite{jap1}, $\gamma_S = (5/2)\gamma_{\text{\tiny $M$}} = 0.8895$
and $K = 0.005685$  \cite{baum}. The parameter $q$ describes the rotation and is  related to the Kerr parameter $a_{\rm s} = S/G_NM^2$ by the relation
\be
q = \frac{C_{\text{\tiny $M$}}^2}{C_{\text{\tiny $J$}}}\left(\frac{M}{C_{\text{\tiny $M$}}}\right)^{-1/2}a_{\rm s}.
\ee
The parameter $\delta_{\rm c}(q)$ is in turn related to the density fluctuations $\delta$ by the relation
\be
\delta = \delta_{\rm c}(q) + \left(\frac{M}{C_{\text{\tiny $M$}}}\right)^{1/\gamma_{\text{\tiny $M$}}} =
 \delta_{0{\rm c}} + K\left(\frac{C_{\text{\tiny $M$}}^2}{C_{\text{\tiny $J$}}}\right)^{2}\left(\frac{M}{C_{\text{\tiny $M$}}}\right)^{-1}a_{\rm s}^2 + \left(\frac{M}{C_{\text{\tiny $M$}}}\right)^{1/\gamma_{\text{\tiny $M$}}},
\ee
such that we can easily go from the variables $\delta$ and $q$ to $M$ and $a_{\rm s}$, respectively.

\begin{figure}[t]
	\includegraphics[width=0.48 \linewidth]{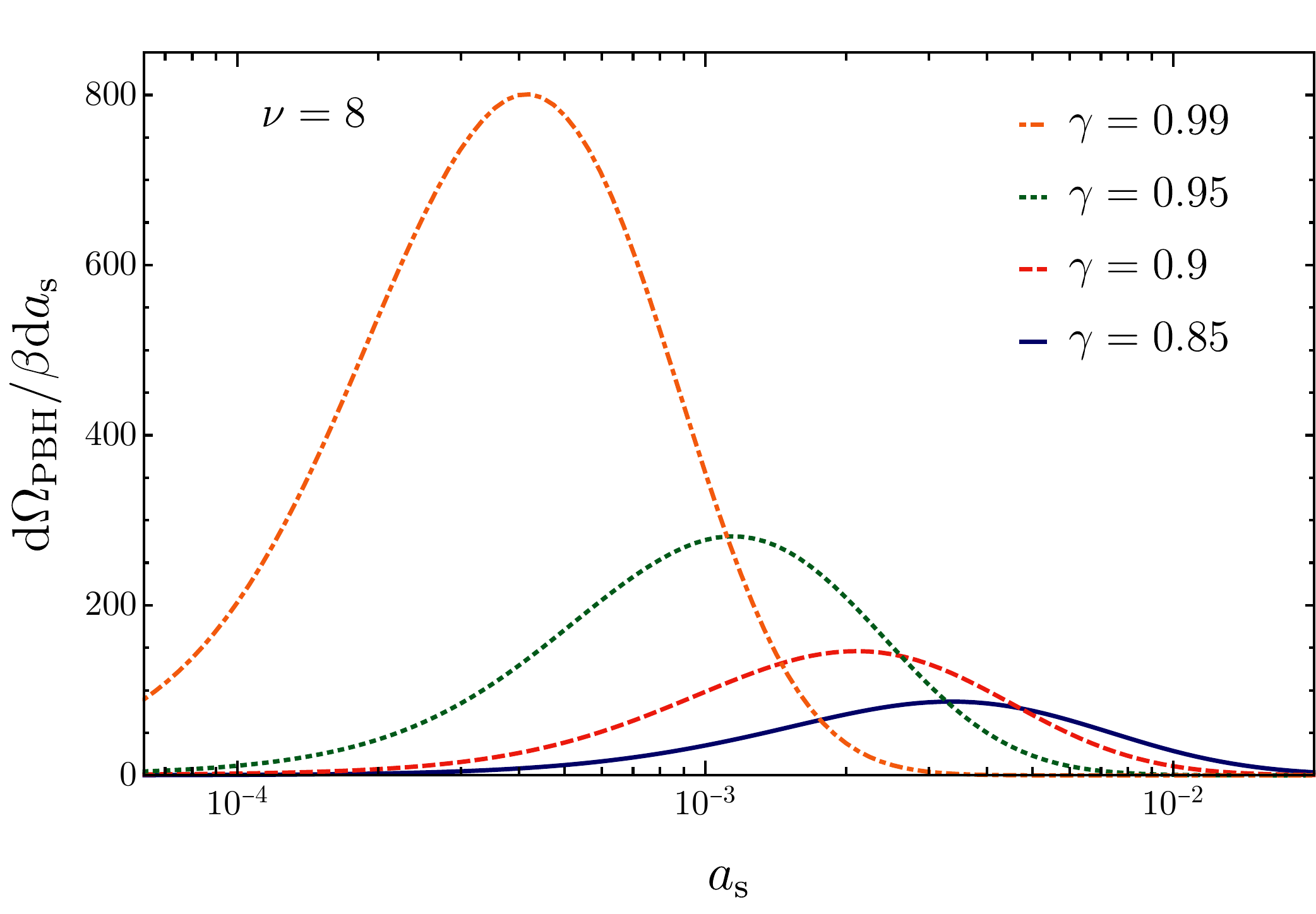}
	\includegraphics[width=0.48 \linewidth]{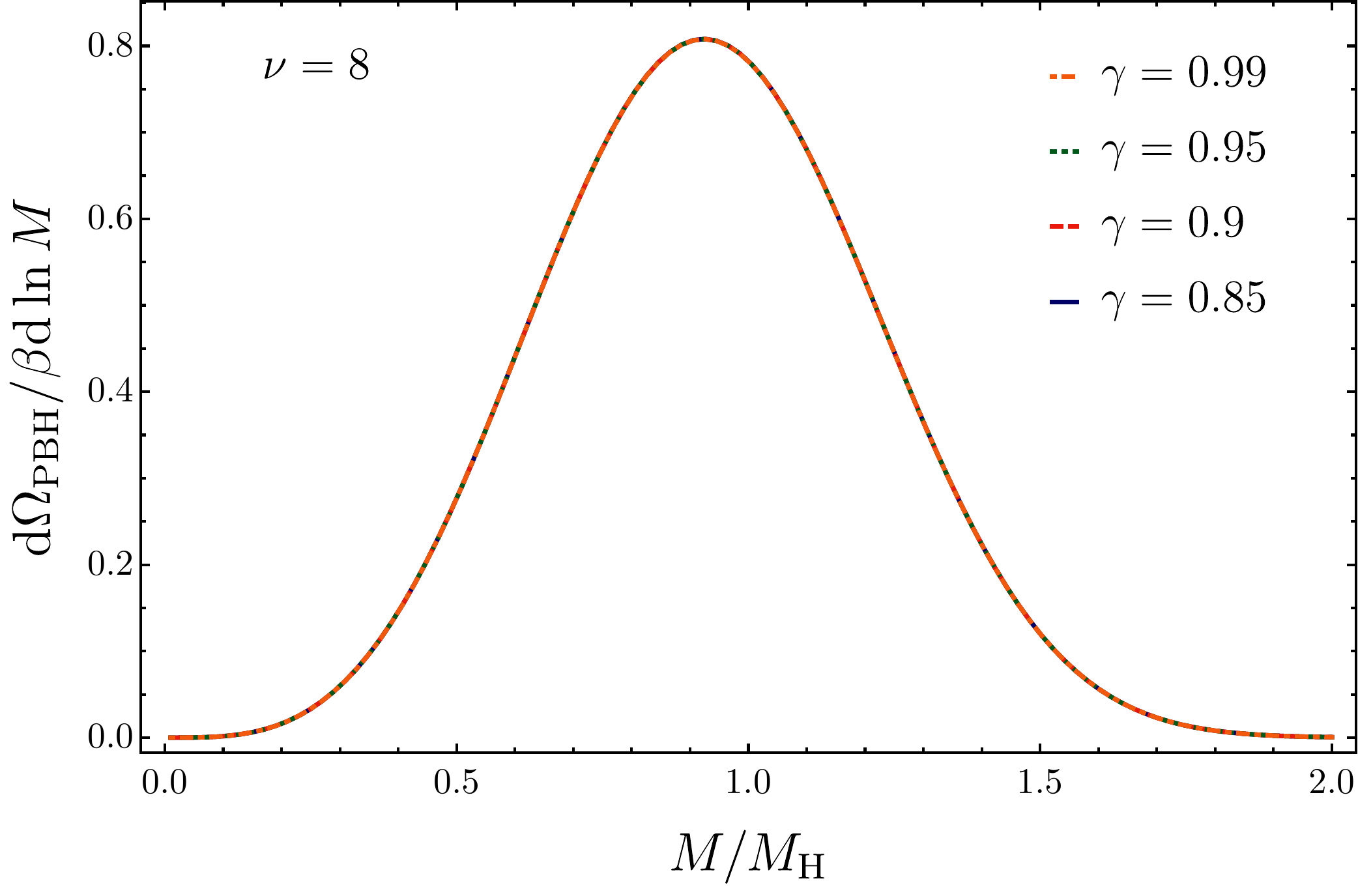}
	\centering
	\caption{ The distributions $\d \Omega_\PBH /\beta \d a_{\rm s}$ and $\d \Omega_\PBH /\beta \d \ln M$ for the distribution $P(a_{\rm s})$.
		The results are obtained with  $C_{\text{\tiny $M$}} = 5.118 M_\H$ and $C_{\text{\tiny $J$}} = 26.19 M_\H$ \cite{Jap}. 
		The distribution of $\d \Omega_\PBH /\beta \d \ln M$ peaks at $M=0.92 M_{\rm H}$.}
	\label{Omega/beta}
\end{figure}

The PBH density parameter  at formation  epoch is given by 
\be
\Omega_{\PBH} = \frac{1}{M_\H}\int_{0}^{\infty}\d q P(q)
 \int_{\delta_{\rm c}(q)}
 \d \delta M(\delta) P(\delta)\,\,\,\,({\rm at}\,\,\,{\rm formation}), 
\ee
where, for demonstrative purposes, we take  $P(\delta)$ to be a Gaussian distribution with variance $\sigma$.
 Since the integration measure is transformed as
\be
\d q \ \d \delta = \frac{C_{\text{\tiny $M$}}}{\gamma_{\text{\tiny $M$}}C_{\text{\tiny $J$}}}\left(\frac{M}{C_{\text{\tiny $M$}}}\right)^{-\frac{3}{2}+\frac{1}{\gamma_{\text{\tiny $M$}}}}\d a_{\rm s} \ \d M, 
\ee
 the density abundance becomes
\be
\label{Omega}
\Omega_{\PBH} = \frac{1}{M_\H}\frac{1}{\sqrt{2\pi}\sigma}\frac{C_{\text{\tiny $M$}}^2}{\gamma_{\text{\tiny $M$}}C_{\text{\tiny $J$}}}\int_{0}^{\infty}\d a_{\rm s} \int\d M \ P(a_{\rm s})\left(\frac{M}{C_{\text{\tiny $M$}}}\right)^{-\frac{1}{2}+\frac{1}{\gamma_{\text{\tiny $M$}}}}e^{-\frac{1}{2\sigma^2}\left(\delta_{0{\rm c}}+ K\left(\frac{C_{\text{\tiny $M$}}^2}{C_{\text{\tiny $J$}}}\right)^{2}\left(\frac{M}{C_{\text{\tiny $M$}}}\right)^{-1}a_{\rm s}^2 + \left(\frac{M}{C_{\text{\tiny $M$}}}\right)^{1/\gamma_{\text{\tiny $M$}}}\right)^2}
\ee
where $P(a_{\rm s})$ is the distribution of the Kerr parameter $a_{\rm s}$ we have previously calculated.
The PBH spin distribution is therefore finally obtained by computing
\be
\frac{\d \Omega_{\PBH}}{\d a_{\rm s}} = \frac{1}{M_\H}\frac{1}{\sqrt{2\pi}\sigma}\frac{C_{\text{\tiny $M$}}^2}{\gamma_{\text{\tiny $M$}}C_{\text{\tiny $J$}}} \int\d M \ P(a_{\rm s})\left(\frac{M}{C_{\text{\tiny $M$}}}\right)^{-\frac{1}{2}+\frac{1}{\gamma_{\text{\tiny $M$}}}}e^{-\frac{1}{2\sigma^2}\left(\delta_{0{\rm c}} + K\left(\frac{C_{\text{\tiny $M$}}^2}{C_{\text{\tiny $J$}}}\right)^{2}\left(\frac{M}{C_{\text{\tiny $M$}}}\right)^{-1}a_{\rm s}^2 + \left(\frac{M}{C_{\text{\tiny $M$}}}\right)^{1/\gamma_{\text{\tiny $M$}}}\right)^2}
\ee
where we have to perform the integration over the masses $M$. 
The amount of PBHs formed at a given epoch by the collapse of a region is described by\footnote{
As we have seen, the generation of a first-order  non-zero spin during the collapse of the overdensity is due to the small deviations from spherical symmetry. One should recall that this has an impact also on the threshold value at which the collapse takes place \cite{kun}.}
\be
\label{beta}
\beta (M) = \frac{1}{\sqrt{2\pi}\sigma}\frac{C_{\text{\tiny $M$}}}{\gamma_{\text{\tiny $M$}} C_{\text{\tiny $J$}}}\int_{0}^{\infty}\d a_{\rm s} \int \d M \ P(a_{\rm s})\left(\frac{M}{C_{\text{\tiny $M$}}}\right)^{-\frac{3}{2}+\frac{1}{\gamma_{\text{\tiny $M$}}}}e^{-\frac{1}{2\sigma^2}\left(\delta_{0{\rm c}} + K\left(\frac{C_{\text{\tiny $M$}}^2}{C_{\text{\tiny $J$}}}\right)^{2}\left(\frac{M}{C_{\text{\tiny $M$}}}\right)^{-1}a_{\rm s}^2 + \left(\frac{M}{C_{\text{\tiny $M$}}}\right)^{1/\gamma_{\text{\tiny $M$}}}\right)^2}.
\ee
In Fig.~\ref{Omega/beta} we plot $\d \Omega /\beta \d a_{\rm s}$ and $\d \Omega /\beta \d \ln M$ for the distribution $P(a_{\rm s})$ of the Kerr parameter
 obtained numerically.
\noindent
\hspace{-0.15cm}
We observe that the Kerr parameter is distributed with a peak which shifts towards smaller values for higher values of $\gamma$, while the impact on the mass distribution is
rather negligible. 
\begin{framed}
{\footnotesize
\noindent 
Our results assume that  PBH spin after the formation does not evolve appreciably. For a  rough estimate we report here the considerations put forward in  Ref. \cite{Jap}.
After formation, the spin of PBH evolves until the present day under the effect of the torque generated by the background radiation fluid. The interaction with the background radiation has the effect of reducing the spin. In particular, one can assess the loss of $S$ as (passing to cosmic time and setting $8\pi G_N$  to unity)
\be
\dot S \sim -M F_{\rm rad}\sim -H^2 M^3 a_{\rm s},
\ee
where we have estimated the force generating the torque as $F_{\rm rad} \sim\overline{ \rho} _{\rm rad} M^2 a_{\rm s}$  \cite{ev1,ev2,Jap} and used the fact that in a radiation dominated universe $H^2 \sim \overline{\rho}_{\rm rad}$.
On the other hand, the PBH mass evolves by accretion from the surrounding radiation fluid as
\be
\dot M \sim \overline{\rho}_{\rm rad} M^2 \sim  H^2 M^2.
\ee
Solving the previous equation tells us that the mass evolution is described by \cite{carr}
\be
M \sim \frac{t}{1+t/t_\H\lp t_\H / M_\H -1\rp }
\ee
where $t_\H$ and $M_\H$ are the initial time and mass, respectively. One can conclude that for PBH with initial masses $M_\H$ smaller than the horizon mass $t_\H$, their accretion is small. Finally, the rate of change of the spin parameter is estimated to be
\be
\dot a_{\rm s} = \frac{\d }{\d t }\lp \frac{S}{M^2}\rp = \frac{\dot S }{M^2} - 2 \frac{S \dot M }{M^3} \sim - H^2 M_\H a_{\rm s},
\ee
resulting in an evolution given by \cite{Jap}
\be
a_{\rm s}\sim a_{\rm s, \H} \exp \llp \alpha  \frac{M_\H}{t_\H} \lp \frac{t_\H}{t}-1\rp\rrp, \qquad \alpha \sim \mathcal{O}\lp1 \rp.
\ee
Therefore, one might expect that the spin parameter evolution can be safely neglected after formation when $M_\H \lsim  t_\H$.  
}
\end{framed}

\section{Conclusions}
In this paper, we have computed the spin probability distribution of PBHs at formation time. We have shown that two ingredients are crucial to generate a spin at first-order in perturbation theory: the collapsing object must be non-spherical, and the velocity shear be misaligned with the inertia tensor of the object. If these conditions are satisfied, a spin of order $10^{-2}$ is generated owing to the action of a first-order tidal torque as the perturbation re-enters the horizon. The spin probability distribution of PBHs is such  that smaller values of the spin are achieved for a collapsing perturbation with higher peak significance and narrower power spectrum.

Our results could be improved in several ways. On the one hand, we have assumed that the initial distributions of the overdensity and velocity are Gaussian. It would be worth assessing what is the impact of a primordial non-Gaussianity by extending our results from peak theory to non-Gaussian distributions. Since PBHs are created from the tails of the probability distribution, we expect this may have an impact on the spin as well.

On the other hand, we have not carefully followed the evolution of the spin after formation. Even though our rough estimates indicate that changes will not be appreciable, 
 it has been argued that spinning PBHs suffer from super-radiant instabilities in the radiation phase (for masses $M \lesssim 0.1 a_{\rm s} M_\odot $) and this may lead to a sizeable reduction of the spin $\delta a_{\rm s} \sim a_{\rm s}$ \cite{pani}. 
 We leave these interesting points for future investigation.

Let us conclude with a brief comparison to the recent literature on the same topic. In Ref. \cite{Jap}, the PBH spin distribution has been derived by integrating the probability $P(s_{\rm e},\delta)$  (or better their proxies $a_{\rm s}$ and $M_\PBH$) over $\delta$ under the assumptions that there is no correlation between
the initial overdensity  and $s_{\rm e}$, and the probability density for $s_{\rm e}$ is flat.  This represents  a limiting case in which the initial overdensity is allowed to have a large angular momentum. Our computations show that the assumption of a flat distribution for $s_{\rm e}$ is not a good starting point.
 
Furthermore, another piece of work, Ref.~\cite{nor}, appeared on the same subject of PBH spin at formation time while we were completing our draft. Even though the  spin probability distribution is not discussed there,    our results coincide
 with theirs as far as the percent level of PBH spin is concerned.  We also agree on  the statement that the PBH spin should vanish 
in the limit of very narrow power spectra, even though for a different reason. Namely, we find that, in this limit, the probability for the off-diagonal elements of the velocity shear is peaked around zero, thus delivering zero spin. This follows from the tendency of the velocity shear to align with the inertia tensor for very narrow power spectra, which inhibits the generation of spin.
By contrast, Ref. \cite{nor} have calculated the spin at second-order in perturbation theory, $a_{\rm s}\sim {\cal R}^2$, in analogy with the argument of \cite{peebles} that the total angular momentum contained in a spherical proto-galaxy  starts at second-order in perturbation theory. As we explained in the introduction, this result is a consequence of the choice of a Lagrangian sphere to describe the collapsing object. When the ellipsoidal shape of the density profile around the peaks is taken into account, the spin can grow at first order for generic power spectra, in analogy with the argument of Refs. \cite{doroshkevich,white}.

\acknowledgments
V. DL., G.F. and A.R. are  supported by the Swiss National Science Foundation (SNSF), project {\sl The Non-Gaussian Universe and Cosmological Symmetries}, project number: 200020-178787.
V.D. acknowledges support by the Israel Science Foundation (grant no. 1395/16).

\appendix

\renewcommand\theequation{\Alph{section}.\arabic{equation}}

\section{Covariance matrix}
\label{appcorr}
In this Appendix we compute the covariant matrix ${\bf M}$, defined in Eq.~\eqref{covm}. All quantities are defined in the CMC gauge and we decide to drop the label to simplify the notation.
Let us start with the computation of the correlator $\langle{\delta^2 } \rangle$.
 We know from the definitions that 
\be
{\delta } (\vx,\eta_\H)&= 
  \frac{V}{(2\pi)^3}\int \d^3k
\, {\delta } (\vk,\eta_\H)
W(k)
  \, e^{i \vec{k} \cdot \vec{x}}, 
  \ee
  thus
\be
\langle {\delta }  (\vec x) {\delta }  (\vec x)\rangle
\equiv 
\langle {\delta^2 } \rangle
= 
 \frac{V}{2\pi^{2}}\int \d k\, k^{2} \,\big\lvert {\delta } (\vk,\eta_\H)\big\lvert^2\,W^2(k) = \sigma_{\delta } ^2.
\ee
Now we can consider all the others. Indeed
\begin{itemize}
	\item $\langle  {\delta } \,   \zeta _{ij}\rangle$:
\be
\langle {\delta } \,   \zeta _{ij}\rangle &= 
V^2
\int \frac{\d ^3 k}{(2 \pi )^3} \int \frac{\d ^3 k'}{(2 \pi )^3} 
\, {\delta } (\vk,\eta_\H) {\delta } (\vk',\eta_\H)
W(k)W(k')
\frac{\partial^2}{\partial y^i \partial y^j}
 e^{i \lp  \vk \cdot \vx+\vk ' \cdot \vy \rp } 
\bigg |_{\vec x \to \vec y}
\\
&=-\frac{V }{2 \pi ^2}
 \int
   \d  k \, 4 \pi k^2  k_i k_j
 \,\big\lvert {\delta } (\vk,\eta_\H)\big\lvert^2
 W^2(k) 
\\
&=-\frac{1}{3}
\delta_{ij}
\frac{V }{2 \pi ^2} 
 \int
  \d  k \,  k^4
 \,\big\lvert {\delta } (\vk,\eta_\H)\big\lvert^2
 W^2(k)
= -\frac{1}{3} \delta_{ij} \sigma_{\times}^2;
\ee
\item $\langle \zeta_i  \zeta_{j}\rangle$:
\be
\langle \zeta_i  \zeta_{j}\rangle &= 
V^2
\int \frac{\d ^3 k}{(2 \pi )^3} \int \frac{\d ^3 k'}{(2 \pi )^3} 
\, {\delta} (\vk,\eta_\H) {\delta} (\vk',\eta_\H)
W(k)W(k')
\frac{\partial^2}{\partial x^i \partial y^j}
 e^{i \lp  \vk \cdot \vx+\vk ' \cdot \vy \rp } 
\bigg |_{\vec x \to \vec y}
\\
&=-\frac{V }{2 \pi^2}
 \int
 \d  k\,  k^2  k_i k_j  
 \,\big\lvert {\delta } (\vk,\eta_\H)\big\lvert^2
 W^2(k)
\\
&=-\frac{1}{3}
\delta_{ij}
\frac{V }{2 \pi ^2} 
 \int
 \d  k\,  k^4
 \,\big\lvert {\delta} (\vk,\eta_\H)\big\lvert^2
 W^2(k)
= -\frac{1}{3} \delta_{ij} \sigma_{\times}^2;
\ee
\item $\langle \zeta_{ij} \zeta_{mn}\rangle$:
\be
\langle \zeta_{ij} \zeta_{mn}\rangle
&=
V^2
\int \frac{\d ^3 k}{(2 \pi )^3} \int \frac{\d ^3 k'}{(2 \pi )^3} 
\, {\delta} (\vk,\eta_\H) {\delta} (\vk',\eta_\H)
W(k)W(k')
\frac{\partial^2}{\partial x^i \partial x^j} \frac{\partial^2}{\partial y^m \partial y^n}
 e^{i \lp  \vk \cdot \vx+\vk ' \cdot \vy \rp } 
\bigg |_{\vec x \to \vec y}
\\
&=\frac{V }{2 \pi ^2}
 \int
 \d  k \,  k^2 k_i k_j k_m k_n  
 \,\big\lvert {\delta} (\vk,\eta_\H)\big\lvert^2
 W^2(k)
\\
& =\frac{1}{15} \lp \delta_{ij} \delta_{mn} +  \delta_{im} \delta_{jn} +\delta_{in} \delta_{jm} \rp 
\frac{V }{2 \pi^2} 
\int
 \d  k  \, k^6
 \,\big\lvert {\delta} (\vk,\eta_\H)\big\lvert^2
 W^2(k)
\\
& = \frac{1}{15} \lp \delta_{ij} \delta_{mn} +  \delta_{im} \delta_{jn} +\delta_{in} \delta_{jm} \rp 
 \sigma_{\zeta}^2;
\ee
therefore one finds $\langle \zeta_{11}\zeta_{11} \rangle  = 3 \langle \zeta_{11}\zeta_{22} \rangle=
3 \langle \zeta_{12}^2 \rangle=...= \sigma _{\zeta} ^2/5$;
\item $\langle \delta\,  \widetilde v_{ij}\rangle$:
\be
\langle {\delta }  \, \widetilde v_{ij}\rangle
&=
 -\frac{ k_\H }{g_v (\eta_\H)} V^2
\int \frac{\d ^3 k}{(2 \pi )^3}\int \frac{\d ^3 k'}{(2 \pi )^3} 
\,\frac{T_v(k',\eta_\H)}{T_\delta(k',\eta_\H)}\, 
\, {\delta } (\vk,\eta_\H) {\delta } (\vk',\eta_\H)
\\
&\times W(k)W(k')
\lp \frac{k_i ' k_j '}{k^{\prime 2}}\rp
 e^{i \lp  \vk \cdot \vx+\vk ' \cdot \vy \rp }
\bigg |_{\vec x \to \vec y} 
\\
&=
 -\frac{ k_\H }{g_v (\eta_\H)} \frac{V }{2 \pi^2}
 \int
 \d  k\,
 k_i k_j
 \,\frac{T_v(k,\eta_\H)}{T_\delta(k,\eta_\H)}\, 
  \,\big\lvert {\delta } (\vk,\eta_\H)\big\lvert^2
  W^2(k)
 \\
 &=-\frac{1}{3} \delta_{ij}
\frac{ k_\H }{g_v (\eta_\H)} \frac{V }{2 \pi ^2}
 \int
  \d  k\, k^2
 \,\frac{T_v(k,\eta_\H)}{T_\delta(k,\eta_\H)}\, 
  \,\big\lvert {\delta } (\vk,\eta_\H)\big\lvert^2
  W^2(k)
 \\
 &\sim-\frac{1}{3} \delta_{ij}
\frac{ k_\H }{g_v (\eta_\H)}  \,\frac{T_v(k_\H,\eta_\H)}{T_\delta(k_\H,\eta_\H)}\,  \frac{V }{2 \pi ^2}
 \int
   \d  k\, k^2
  \,\big\lvert {\delta } (\vk,\eta_\H)\big\lvert^2
  W^2(k)
=
  -\frac{1}{3}\delta_{ij} \sigma_{\delta };
\ee

\item $\langle \zeta_{ij} \widetilde v_{mn}\rangle$:
 \be
\langle \zeta_{ij} \widetilde v_{mn}\rangle
&=
 -\frac{ k_\H }{g_v (\eta_\H)} V^2
\int \frac{\d ^3 k}{(2 \pi )^3}\int \frac{\d ^3 k'}{(2 \pi )^3} 
\,\frac{T_v(k',\eta_\H)}{T_\delta(k',\eta_\H)}\, 
\, {\delta } (\vk,\eta_\H) {\delta } (\vk',\eta_\H)
\\
&\times W(k)W(k')
\lp \frac{k_m ' k_n '}{k^{\prime 2}}\rp
\frac{\partial^2}{\partial x^i \partial x^j}
 e^{i \lp  \vk \cdot \vx+\vk ' \cdot \vy \rp }
\bigg |_{\vec x \to \vec y} 
\\
&=
 \frac{ k_\H }{g_v (\eta_\H)} \frac{V }{2 \pi ^2}
 \int
 \d  k \,
k_i k_j k_m k_n
 \,\frac{T_v(k,\eta_\H)}{T_\delta(k,\eta_\H)}\, 
  \,\big\lvert {\delta } (\vk,\eta_\H)\big\lvert^2
  W^2(k)
 \\
 &=\frac{1}{15} \lp \delta_{ij} \delta_{mn} +  \delta_{im} \delta_{jn} +\delta_{in} \delta_{jm} \rp 
\frac{ k_\H }{g_v (\eta_\H)} \frac{V }{2 \pi ^2}
 \int
 \d  k \,
k^4
 \,\frac{T_v(k,\eta_\H)}{T_\delta(k,\eta_\H)}\, 
  \,\big\lvert {\delta } (\vk,\eta_\H)\big\lvert^2
  W^2(k) 
 \\
 &\sim\frac{1}{15} \lp \delta_{ij} \delta_{mn} +  \delta_{im} \delta_{jn} +\delta_{in} \delta_{jm} \rp
\frac{ k_\H }{g_v (\eta_\H)}  \,\frac{T_v(k_\H,\eta_\H)}{T_\delta(k_\H,\eta_\H)}\,  \frac{V }{2 \pi ^2}
 \int
  \d  k \,
k^4
  \,\big\lvert {\delta } (\vk,\eta_\H)\big\lvert^2
  W^2(k)
 \\
 &=
  \frac{1}{15} \lp \delta_{ij} \delta_{mn} +  \delta_{im} \delta_{jn} +\delta_{in} \delta_{jm} \rp \frac{\sigma_{\times}^2}{\sigma_{\delta }};
  \ee
therefore one finds 
$\langle \zeta_{11} \widetilde v_{11} \rangle  =
 3 \langle \zeta_{11} \widetilde v_{22} \rangle=
3 \langle \zeta_{12} \widetilde v_{12} \rangle=...=  \sigma_{\times}^2 / 5 \sigma_{\delta }$;

\item $\langle \widetilde v_{ij} \widetilde v_{mn}\rangle$:
\be
\langle \widetilde v_{ij} \widetilde v_{mn}\rangle
&=
 \lp \frac{ k_\H }{g_v (\eta_\H)}\rp ^2 V^2
\int \frac{\d ^3 k}{(2 \pi )^3}\int \frac{\d ^3 k'}{(2 \pi )^3} 
\,\frac{T_v(k',\eta_\H)}{T_\delta(k',\eta_\H)}\, 
\frac{T_v(k,\eta_\H)}{T_\delta(k,\eta_\H)}\, 
\, {\delta } (\vk,\eta_\H) {\delta } (\vk',\eta_\H)
\\
&\times W(k)W(k')
\lp \frac{k_i k_j}{k^{ 2}}\rp
\lp \frac{k_m ' k_n '}{k^{\prime 2}}\rp
e^{i \lp  \vk \cdot \vx+\vk ' \cdot \vy \rp }
\bigg |_{\vec x \to \vec y} 
\\
&=
 \lp \frac{ k_\H }{g_v (\eta_\H)}\rp ^2 \frac{V }{2 \pi ^2}
 \int
\d  k  \, \frac{k_i k_j k_m k_n}{k^2} 
 \, \frac{T_v^2(k,\eta_\H)}{T^2_\delta(k,\eta_\H)} \, 
  \,\big\lvert {\delta } (\vk,\eta_\H)\big\lvert^2
  W^2(k)
 \\
 &=\frac{1}{15} \lp \delta_{ij} \delta_{mn} +  {\rm cycl.} \rp 
 \lp \frac{ k_\H }{g_v (\eta_\H)}\rp ^2 \frac{V }{2 \pi ^2}
 \int
\d k\, k^2
\, \frac{T_v^2(k,\eta_\H)}{T^2_\delta(k,\eta_\H)} \, 
  \,\big\lvert {\delta } (\vk,\eta_\H)\big\lvert^2
  W^2(k)
 \\
 &\sim \frac{1}{15} \lp \delta_{ij} \delta_{mn} + {\rm cycl.} \rp
 \lp \frac{ k_\H }{g_v (\eta_\H)}\rp ^2  
\, \frac{T_v^2(k_\H,\eta_\H)}{T^2_\delta(k_\H,\eta_\H)} \, 
 \frac{V }{2 \pi ^2}
 \int
 \d k\, k^2
  \,\big\lvert {\delta } (\vk,\eta_\H)\big\lvert^2
  W^2(k)
 \\
 &=
  \frac{1}{15} \lp \delta_{ij} \delta_{mn} +  \delta_{im} \delta_{jn} +\delta_{in} \delta_{jm} \rp;
  \ee
therefore one finds $\langle  \widetilde v_{11}^2 \rangle  = 3 \langle  \widetilde  v_{11} \widetilde v_{22} \rangle=
3 \langle  \widetilde v_{12}^2 \rangle= ...= 1/5$.
\end{itemize}

\end{document}